\documentclass{lmcs}

\usepackage{amsmath}
\usepackage{amsfonts}
\usepackage{amssymb}
\usepackage{amsthm}
\usepackage{graphicx}
\usepackage{amscd}
\usepackage{latexsym}
\usepackage{hyperref}
\usepackage{tikz}
\usepackage{tikz-cd}
\usepackage{color}
\usepackage{MnSymbol}

\usepackage[colorinlistoftodos]{todonotes}
\usepackage{changes}

\newcommand{\Pow}{\mathsf{P}}
\newcommand{\dom}{d^-}
\newcommand{\cod}{d^+}
\newcommand{\Dom}{D^-}
\newcommand{\Cod}{D^+}

\newcommand{\id}{\mathit{id}}

\begin{document}
\title{Generalised Möbius Categories and Convolution Kleene Algebras}

\author[J.~Cranch]{James Cranch\lmcsorcid{0009-0003-1307-8552}}[a]
\author[G.~Struth]{Georg Struth\lmcsorcid{0000-0001-9466-7815}}[a]
\author[J.~Wagemaker]{Jana Wagemaker\lmcsorcid{0000-0002-8616-3905}}[b]

\address{University of Sheffield}
\email{j.d.cranch@sheffield.ac.uk, g.struth@sheffield.ac.uk}
\address{Radboud University}
\email{j.wagemaker@ru.nl}

\begin{abstract}
  Convolution algebras on maps from structures such as monoids, groups
  or categories into semirings, rings or fields abound in mathematics
  and the sciences. Of special interest in computing are convolution
  algebras based on variants of Kleene algebras, which are additively
  idempotent semirings equipped with a Kleene star.  Yet an obstacle
  to the construction of convolution Kleene algebras on a wide class
  of structures has so far been the definition of a suitable star. We
  show that a generalisation of Möbius categories combined with a
  generalisation of a classical recursive definition of a star for
  formal power series by Kuich and Salomaa allow such a
  construction. We discuss several instances of this construction on
  generalised Möbius categories: convolution Kleene algebras with
  tests, modal convolution Kleene algebras, concurrent convolution
  Kleene algebras and higher convolution Kleene algebras (e.g. on
  strict higher categories and higher relational monoids). These are
  relevant to the verification of weighted and probabilistic
  sequential and concurrent programs, using quantitative Hoare logics
  or predicate transformer algebras, as well as to algebraic reasoning
  in higher-dimensional rewriting. We also adapt the convolution
  Kleene algebra construction to Conway semirings, which is widely
  studied in the context of weighted automata, and to star-continuous
  Kleene algebras. Finally, we outline some limitations of this
  construction, compare the convolution Kleene algebra construction
  with a previous construction of convolution quantales and present
  concrete example structures in preparation for future applications.

  \vspace{\baselineskip}


\vspace{\baselineskip}


\end{abstract}

\maketitle

%%%%%%%%%%%%%%%%%%%%%%%%%%%%%%%%%%%%%%%%%%%%%%%%%%%%%%%%%%%%%%%%%%%%%%%%%%%

\section{Introduction}\label{s:introduction}

This article is part of a series on convolution algebras and their
applications in
computing~\cite{BanniserHS21,CranchDS21,DongolHS16,DongolHS21,FahrenbergJSZ23,CalkMPS25}. It
is influenced by work of Schützenberger on formal power
series~\cite{DrosteK09}, Rota on incidence
algebras~\cite{Rota64,JoniR79}, Cartier and Foata on combinatorics on
words~\cite{CartierF69}, and Jónsson and Tarski on boolean algebras
with operators~\cite{JonssonT51}. Also related are group and category
algebras in representation theory and categorical approaches to
topology. Applications of such convolution algebras range from path
algorithms~\cite{AhoHU75,Mehlhorn84,Mohri02}, network
protocols~\cite{Sobrinho03,GriffinG08} and speech
recognition~\cite{MohriPR02} via fuzzy sets and
relations~\cite{Goguen67,EklundGHK18} and provenance analysis and
semiring semantics for logics and games~\cite{GraedelT20}, to
probabilistic and weighted
programming~\cite{LairdMMP13,BelleR20,BatzGKKW22} and
rewriting~\cite{BournezK02,Faggian22,GavazzoF23}, and to quantitative
program verification~\cite{BatzKKMN19,Haslbeck21,FahrenbergJSZ23}, for
a few indicative references.

In these situations one considers maps from a structure $X$, typically
a (co)mo\-noid, category or relational structure, into an algebra $V$ of values, probabilities, weights or costs, such as 
a semiring, ring, field or quantale. The algebra on the function space $V^X$ often
carries the same structure as $V$; it extends to a module structure
when elements in $V^X$ are multiplied with those in $V$.

In work on formal power series or combinatorics on words, $X$ is
typically a monoid and in particular the free monoid generated by a
finite alphabet. In work on incidence algebras, $X$ is a set of
(closed) intervals on a locally finite poset; for path algorithms, $X$
is a set of finite paths on a directed graph. All these examples are
subsumed when $X$ is a category, while applications such as weighted
shuffle languages require a generalisation of $X$ to a set equipped
with a ternary relation satisfying the laws of relational monoids
(monoid objects in the category
$\mathbf{Rel}$~\cite{Rosenthal97,KenneyP11}), of which categories are
special cases.

The considerable interest of semirings and Kleene algebras in
applications led us to study maps $C\to K$ from a category or relational
monoid $C$ into a Kleene algebra, an additively idempotent semiring
$(K,+,\cdot,0,1)$ equipped with a Kleene star
$(-)^\ast:K\to K$. Formally, a Kleene algebra
is an additively idempotent semiring $(S,+,\cdot,0,1)$ with a star
operation $(-)^\ast:S\to S$ satisfying
\begin{equation*}
  1 +\alpha\cdot \alpha^\ast = \alpha^\ast,\qquad \gamma + \alpha\cdot
  \beta \le \beta \implies \alpha^\ast \cdot \gamma \le \beta,
\end{equation*}
and  dual axioms obtained by intechanging the factors in
multiplications~\cite{Conway71,Kozen94}. Here, $\le$ is the partial order defined by
$\alpha\le\beta \iff \alpha+\beta=\beta$, as in any
semilattice.

We wish to construct a Kleene algebra on $K^{C}$ and describe the
situation for a category $C$.  As usual, addition extends pointwise
from $C$ and $K$ to $K^{C}$. The additive unit on $K^{C}$ is the
constant zero map and the multiplicative unit on $K^{C}$ the indicator
map for identity morphisms in $C$, mapping to $0$ and $1$ in $K$. The
multiplication of $f,g:C\to K$ is the convolution
\begin{equation*}
  (f \ast g)(x) = \sum_{x=y\odot z} f(y)\cdot g(z),
\end{equation*}
where $x$, $y$, $z$ are arrows in $C$, $\odot$ is arrow composition,
$\cdot$ is multiplication in $K$ and $\sum$ indicates a finite sup in
$K$.  The decomposition of $x$ into $y$ and $z$ in $C$ must thus be
controlled to make sups finite and thus expressible in $K$. This can
be achieved by restricting to finitely supported functions in $K^{C}$
or to finite value algebras $K$. Yet applications seem to benefit in
particular from a wide class of categories $C$. In the tradition of
Rota's incidence algebras, we thus assume that all arrows in $C$ have
finitely many $2$-decompositions.

The missing piece in the construction of convolution Kleene algebras
has so far been the Kleene star on $K^C$. We generalise Kuich and
Salomaa's classical recursive definition for formal power
series~\cite{KuichS86}. For the free monoid $A^\ast$ on an alphabet
$A$ and a Kleene algebra $K$, specifically, the Kleene star of
$f:A^\ast\to K$ is defined, for the empty word $\varepsilon$ and
$w\neq \varepsilon$, as $f^\ast (\varepsilon)= f(\varepsilon)^\ast$
and
$f^\ast (w) = f(\varepsilon)^\ast \cdot \sum_{w=uv, u\neq \varepsilon}
f(u)\cdot f^\ast (v)$, where the sum ranges over $u$ and
$v$. Generalising $A^\ast$ to a category $C$ we define for each
$f:C\to K$, identity arrow $e$ in $C$ and non-identity arrow $x$ in
$C$, the Kleene star $f^\ast:C\to K$ as
\begin{equation*}
  f^\ast (e)= f(e)^\ast, \qquad f^\ast (x) = f(s(x))^\ast \cdot \sum_{x= y \odot z, y\neq s(x)} f(y)\cdot f^\ast (z),
\end{equation*}
where $s(x)$ denotes the source of the arrow $x$, represented as an
identity arrow, and the sum ranges over $y$ and $z$. Once again, $C$
must be finitely $2$-decomposable to obtain finite sums. Yet for a
recursive definition that supports inductive proofs of the star axioms
of Kleene algebras in $K^{C}$ it seems desirable to associate a notion
of length with each arrow of $C$, which ensures that
$\ell(x)\ge \ell(y)+\ell(z)$ whenever $x=y\odot z$.

Category theory supplies both features: $\ell$-categories equip all
morphisms with a notion of length~\cite{Mitchell72}; Möbius categories
are $\ell$-categories in which all morphisms are finitely
$2$-decomposable~\cite{Leroux75}. 

Our main conceptual contribution consists in showing that the Möbius
condition is precisely what we need to make Kuich and Salomaa's Kleene
star work with categories and relational monoids; our main technical
result (\autoref{thm:main-theorem}) shows that the convolution algebra of
$K$-valued maps on a Möbius category (or relational Möbius monoid) forms
a Kleene algebra.

The Möbius conditions guarantee that the sum over all decompositions
of $x$ remains finite, that the star axioms on $K^C$ can be verified
by induction on $\ell(x)$, and that the proof works for categories and
relational monoids $C$
with multiple objects as in intervals over posets or in path
categories on graphs.  It generalises previous constructions with
value algebras similar to Kleene algebras, where only structures $C$ with
a single identity element could be dealt with (cf.~\cite{CranchDS21}).

The recursive Kleene star is indeed natural for applications. The
second equation unfolds into
\begin{equation*}
  f^\ast(x) = \sum_{1\le i \le \ell(x)}\sum_{x= x_1 x_2 \dots x_i}
  \left(
\prod_{1\le j\le i} f(s(x_j))^\ast\cdot f(x_j)
\right) \cdot f(t_i(x_i))^\ast,
\end{equation*}
where the inner summation ranges over $x_1,\dots,x_i$ and $\ell(x)$
indicates the length of the non-identity arrow $x$. Hence $f^\ast(x)$
maps $f$ over the non-identity arrows in any decomposition of $x$,
interleaved with $f^\ast$ on the objects connecting these arrows (or
on the corresponding identity arrows); it multiplies the weights along
chains of arrows and chooses the ``best'' among the weights of these
decompositions. Similar constructions are familiar from path
algorithms or dynamic programming. Yet note that the notion of length
needed for this recursive definition means that this star is not
suitable for weighted relations or matrices, where the underlying
category (the pair groupoid) lacks a non-trivial notion of length. The
star must then be defined by other means, as explained in
\autoref{ex:conv-kas}.

To demonstrate the versatility of the general construction of the
Kleene star in \autoref{thm:main-theorem}, we provide variants for
$\ast$-continuous Kleene algebras~\cite{Kozen94}
(\autoref{thm:star-cont-ka-star}) and Conway
semirings~\cite{Conway71,BloomE93} (\autoref{thm:conway-semirings}). The
latter are semirings with equational star axioms that feature
prominently in the formal power series literature~\cite{DrosteK09}.
As a contextualisation, we derive the recursive star axioms in
convolution quantales based on arbitrary relational
monoids (\autoref{prop:conv-star}), using the standard Kleene star for
quantales formalised as a sup of powers. Generalising an observation
by Sedlár~\cite{Sedlar24}, we also show (\autoref{thm:main-theorem-kat})
that any convolution Kleene algebra constructed along the lines of
\autoref{thm:main-theorem} forms a Kleene algebra with tests, a formalism
popular for program verification~\cite{Kozen97}.

Beyond these fundamental results, \autoref{thm:main-theorem} allows
constructing convolution Kleene algebras on arbitrary Möbius
categories, where previously only convolution quantales could be
obtained, a more restricted class. This works for suitable relational
monoids, but we conceal them here to keep explanations simple.
\begin{itemize}
\item The interval temporal logics~\cite{Moszkowski12} and duration
  calculi~\cite{ZhouH04} used in program verification ressemble
  incidence Kleene algebras on categories of closed intervals on
  linear orders -- categories with many objects. Previously, semantics
  for quantitative variants have been formalised via incidence
  quantales~\cite{DongolHS21}. Now, \autoref{thm:main-theorem} supplies
  the missing Kleene star -- the chop-star of interval temporal logic
  -- for more faithful semantics based on incidence Kleene algebras
  (\autoref{ex:conv-kas}).

\item Quantitative predicate transformer algebras, dynamic logics and
  Hoare logics have previously been formalised via modal convolution
  quantales on arbitrary categories~\cite{FahrenbergJSZ23}. In
  \autoref{s:modal-convolution-ka} we use \autoref{thm:main-theorem} to
  model them as modal convolution Kleene algebras~\cite{DesharnaisS11}
  on Möbius categories (\autoref{cor:modal-ka1} and \autoref{cor:modal-ka}),
  yet need to restrict function spaces to construct modal
  operators. 

\item Concurrent Kleene algebras and quantales provide algebraic
  interleaving and partial-order semantics for concurrent programs and
  systems~\cite{HoareMSW11}. Quantitative variants have so far been
  formalised as interchange convolution quantales on arbitrary strict
  $2$-categories~\cite{CranchDS21}.  In
  \autoref{s:concurrent-convolution-ka} we adapt \autoref{thm:main-theorem}
  to construct interchange convolution Kleene algebras on Möbius
  $2$-categories (\autoref{cor:ic-ka}), supplying a sequential and a
  concurrent Kleene star for this setting.

\item In~\cite{CalkMPS25}, the constructions of modal and interchange
  convolution quantales mentioned have been combined into a
  construction of convolution $n$-quantales on strict
  $n$-categories. Instead of such higher quantales, (convolution)
  $n$-Kleene algebras have been proposed previously
  in~\cite{CalkGMS22} for applications in higher-dimensional
  rewriting~\cite{AraBGMMM23}. In \autoref{s:convolution-n-ka} we
  describe their construction on Möbius $n$-categories
  (\autoref{thm:n-ka}), adding once again the $n$ Kleene stars required. 
\end{itemize}

The constructions of convolution Kleene algebras focus on the arrows
of categories. Single-set categories~\cite[Chapter XII]{MacLane98}
 simplify their formalisation. For similar reasons we present
relational monoids as catoids $(C,\odot,s,t)$ formed by a set $C$, a
set-valued operation $\odot:C\times C\to \Pow C$, replacing the
ternary relation $C\times C\times C\to 2$ as the multiplication of the
relational monoid, and source an target maps $s,t:C\to C$ as in
(single-set) categories, replacing the unit set of the relational
monoid~\cite{FahrenbergJSZ23}. We recall their properties in
\autoref{s:catoids}. Möbius catoids, a generalisation of Möbius
categories, are introduced in \autoref{s:moebius-catoid}, Möbius
$2$-catoids and Möbius $n$-catoids in
\autoref{s:concurrent-convolution-ka} and \autoref{s:convolution-n-ka}.

For convolution quantales, the results
in~\cite{CranchDS21,FahrenbergJSZ23,CalkMPS25} yield not only extensions from catoids and value quantuales to convolution quantales but correspondences between these structures. Equations in convolution and value quantales also induce equations in the underlying catoid. In order to construct the modal,
concurrent and higher quantales mentioned above,
it is therefore necessary and sufficient to produce a suitable catoid
or category, which is often a much simpler object. 
These results generalise to Kleene algebras except that a suitable on the convolution algebra needs to be defined separately. 

Sections~\ref{s:convolution-kas}--\ref{s:convolution-n-ka},
Section~\ref{s:star-quantale} and the conclusion feature comparisons
of convolution quantales and convolution Kleene algebras. While the
former can be defined on arbitrary catoids, the latter require Möbius
catoids, which still cover many interesting applications. Kleene
algebras seem more appropriate than quantales for algebras defined by
generators and relations à la rational power series, or for program
verification, where programs are finitary constructions generated by
atomic tests and commands, so that infinite nondeterministic choices
supported by quantales are not implementable.

%%%%%%%%%%%%%%%%%%%%%%%%%%%%%%%%%%%%%%%%%%%%%%%%%%%%%%%%%%%

\section{Catoids}\label{s:catoids}

Catoids~\cite{FahrenbergJSZ23} are relational monoids -- monoid
objects in $\mathbf{Rel}$~\cite{Rosenthal97,KenneyP11} -- in algebraic
form.  They are also simple generalisations of categories, single-set
categories~\cite[Chapter XII]{MacLane98} to be precise. Here we
outline their basic properties.

A \emph{catoid} $(C,\odot,s,t)$ consists of a set $C$, a set-valued
operation $\odot:C\times C\to \Pow C$ and source and target maps
$s,t:C\to C$ such that, for all $x,y,z\in C$,
\begin{equation*}
  \bigcup_{v \in y\odot z}x \odot v = \bigcup_{u\in x\odot y} u\odot
  z,\qquad
  x\odot y \neq \emptyset \implies t(x)=s(y),\qquad s(x)\odot x =
  \{x\},\qquad x \odot t(x)=\{x\}.
\end{equation*}
The first axiom is an associativity law. Extending $\odot$ to
$\Pow C\times \Pow C\to \Pow C$ as
$X\odot Y= \bigcup_{x\in X, y\in Y} x \odot y$ and dropping some set
braces allows rewriting it as $x\odot (y\odot z)= (x\odot y)\odot
z$. The second axiom is reminiscent of the definedness condition of
arrow composition in categories: we call $x$ and $y$ are
\emph{composable} if $x\odot y\neq \emptyset$. The third and fourth
catoid axioms are left and right identity axioms similar to those for
categories.

A \emph{category} is a local functional catoid, where a catoid $C$ is
\emph{local} if $t(x)=s(y)$ implies that $x$ and $y$ are composable,
for all $x,y\in C$, and \emph{functional} if $x,x'\in y\odot z$ imply
$x=x'$ for all $x,x',y,z\in C$.

In every functional catoid $C$, $\odot$ specialises to a partial
operation on $C$, which maps each composable pair $(x,y)$ of elements
to the unique $z\in x\odot y$ and is undefined otherwise. When working
with categories, we often use this partial operation tacitly to avoid
set braces.

The laws in the following lemma are used for calculating with
catoids below; see~\cite{FahrenbergJSZ23} for details.

\begin{lem}\label{lem:catoid-props}
  In every catoid,
  \begin{enumerate}
  \item \label{item:double} $s\circ s= s$, $t\circ t=t$, $s\circ t=t$ and $t\circ s=s$,
  \item $s(x)= x$ if and only $t(x)=x$,
  \item $s(x)\odot s(x)=\{s(x)\}$ and $t(x)\odot t(x)=\{t(x)\}$,
  \item $s(x)\odot t(y)=t(y)\odot s(x)$,
   \item $s(s(x)\odot y)=s(x)\odot s(y)$ and $t(x\odot
     t(y))=t(x)\odot t(y)$,
      \item $s(x\odot y)\subseteq s(x\odot s(y))$ and $t(x\odot
    y)\subseteq t(t(x)\odot y)$,
  \item $s(x\odot y) = \{s(x)\}$ and $t(x\odot y) = \{t(y)\}$ if $x$,
    $y$ are composable,
 \item $x\in y\odot z$ implies $s(x)=s(y)$ and $t(x)=t(z)$.
  \end{enumerate}
\end{lem}

Some of these laws are related by \emph{opposition}. As for
categories, this means exchanging the arguments in compositions as
well as source and target maps. The class of catoids is closed under
opposition; the opposite of each theorem about catoids is a theorem.

\autoref{lem:catoid-props}(2) implies that the set of fixpoints of
$s$ equals the set of fixpoints of $t$.  We write $C_0$ for this set
and refer to its elements as \emph{identities} of $C$. In a category,
identities are identity arrows, and thus in one-to-one correspondence
with objects. We also write $C_1=C-C_0$ for the set of
``nondegenerate'' elements of $C$. It is easy to show using (1) that
$C_0$ is equal also to $s(C)$ and $t(C)$, the image of $C$ under $s$
and $t$, respectively. Further, elements of $C_0$ are orthogonal
idempotents:

\begin{lem}\label{lem:catoid-orth-idem}
  Let $C$ be a catoid. Then, for all $x,y\in C_0$,
  \begin{equation*}
    x\odot y =
    \begin{cases}
      \{x\} & \text{ if } x=y,\\
      \emptyset & \text{ otherwise}.
    \end{cases}
  \end{equation*}
\end{lem}

\begin{exa}\label{ex:catoids}
  The following catoids appear across this text.
  \begin{enumerate}
  \item The free monoid $(A^\ast,\cdot,\varepsilon)$ on the set $A$,
    which forms a category.
  \item The \emph{shuffle catoid} $(A^\ast,\|,\varepsilon)$ on $A$
    with the shuffle multioperation
    $\|: A^\ast \times A^\ast \to \Pow A^\ast$  defined, for all
    $a,b\in \Sigma$ and $v,w\in \Sigma^\ast$, by
    \begin{equation*}
      v\parallel \varepsilon = \{v\}=\varepsilon \parallel
      v\qquad\text{and}\qquad
      (av)\parallel (bw) = a (v\parallel (bw))\cup b ((av)\parallel w)
      \end{equation*}
    forms a catoid, but not a category.
  \item The \emph{interval category} on the poset $(P,\le)$ is simply
    this poset viewed as a category (with at most one arrow in each
    homset). As a single-set category we write $(I_P,\odot,s,t)$,
    where $I_P$ is the set of closed intervals on $P$, and where
    interval composition $\odot:I_P\times I_P\to I_P$ and maps
    $s,t:I_P\to I_P$ are given by
  \begin{equation*}
    [a,b]\odot [c,d]=
    \begin{cases}
      [a,d]& \text{ if } b=c,\\
      \text{undefined} & \text{ otherwise},
    \end{cases}
\qquad s([a,b])= [a,a],\qquad t([a,b])=[b,b].
\end{equation*}

\item 
    The \emph{pair groupoid} $(X\times X,\odot,\{(a,a)\}_{a\in X})$ on
  the set $X$ is $X\times X$ viewed as a poset. As a single-set-category it has
  composition $\odot:(X\times X) \times (X\times X)\to X\times X$ defined as in (3), replacing intervals with ordered
  pairs (inverses are ignored). 

\item The \emph{path category} $P(G)$ on a directed graph $G$ is
  the free category generated by $G$~\cite{MacLane98}.

\item The \emph{guarded strings}~\cite{KozenS97} on the sets $T$ and
  $A$ form a path-like category $P(T,A)$ generated by the elements in
  $T$ and $A$. Atomic guarded strings are formed by elements of $T$ or
  paths of the form $(t,a,t')$ where $t,t'\in T$ and $a\in A$. The set
  of guarded strings is the smallest set containing the atoms and
  closed under path composition, defined as in (5), assuming that
  different atoms do not compose. Sources and targets of paths are the
  elements in $T$ at the beginning and end of paths. Elements in $T$
  thus form the units in this category, elements $(t,a,t')$ correspond
  to non-degenerate edges.
\end{enumerate}
\end{exa}

\begin{rem}\label{rem:quiver}
  In \autoref{ex:catoids}(5), we may model a directed graph as a
  set $G$ (of edges) equipped with source and target maps $s,t:G\to G$
  satisfying $s\circ s=s=t\circ s$ and $t\circ t= t=s\circ t$.  The set $G_0$ (of vertices) is then given by the set of
  fixpoints of $s$, which equals the set of fixpoints of $t$, viewing
  vertices as degenerate edges.  A path can be modelled either as an
  element of $G_0$ or a sequence $(x_0,\dots x_n)$ of nondegenerate
  edges in $G$ in which $t(x_i) =s(x_{i+1})$ for all $0\le
  i<n$. Source and target maps on paths take the source and target of
  the first and last element in a given path, respectively. Path
  composition is $\pi_1\odot \pi_2$ is $\pi_1\pi_2$ if
  $t(\pi_1)=s(\pi_2)$ and undefined otherwise. The identities of path
  composition are the constant paths, the elements of $G_0$.
\end{rem}

\begin{rem}
  Catoids can be modelled alternatively as relational structures with
  a ternary relation because
  $C\times C\to \Pow C \simeq C\times C\times C\to 2$. The catoid
  axioms translate into the laws of \emph{relational monoids}, which
  are monoid objects in the monoidal category
  $\mathbf{Rel}$. Multiplication in a relational monoid is a ternary
  relation and its identity a set, which can be modelled as the set of
  fixpoints of the source and target maps; see~\cite{FahrenbergJSZ23} for
  details.  % Formally,
\end{rem}

%%%%%%%%%%%%%%%%%%%%%%%%%%%%%%%%%%%%%%%%%%%%%%%%%%%%%%%%%

\section{Möbius catoids}\label{s:moebius-catoid}

Möbius categories have been proposed by Leroux and
colleagues~\cite{Leroux75,ContentLL80,Leroux82} as a uniform framework
for the Möbius functions used in the combinatorics on words by Cartier
and Foata~\cite{CartierF69} and the foundations of combinatorics by
Rota~\cite{Rota64,JoniR79}. Their essential properties have also been
summarised by Lawvere and Menni~\cite{LawvereM10}. We generalise to
Möbius catoids, where their most important properties still hold.

Let $C$ be a catoid.
\begin{itemize}
\item A \emph{decomposition of degree $n$} (an
  \emph{$n$-decomposition}) of an element $x\in C$ is a finite list
  $(x_1,\dots,x_n)$, $n\ge 0$, of non-identity elements in $C$ such
  that $x\in x_1 \odot \dots \odot x_n$.
\item An element $x\in C$ is \emph{indecomposable} if it has no
  $n$-decomposition for $n > 1$.
\item The \emph{length} $\ell(x)$ of an element $x\in C$ is the sup of
  its degrees of decomposition, with $\infty$ assigned if there is no
  finite sup.
\item An element of $C$ is \emph{finitely $n$-decomposable} if it has finitely
  many $n$-decompositions.
\item An element of $C$ is \emph{finitely decomposable} if it has
  finitely many decompositions.
\end{itemize}

An \emph{$\ell$-catoid} is a catoid in which each element has finite
length. A \emph{Möbius catoid} is a catoid in which each element is
finitely decomposable. A catoid is \emph{finitely $2$-decomposable} if
every element has this property.

By definition, an element $x$ of a catoid is finitely decomposable if
and only if $\ell(x) < \infty$ and it is finitely $n$-decomposable for
each $n\le \ell(x)$. A Möbius catoid is thus an $\ell$-catoid in which
each element $x$ is finitely $n$-decomposable for each $n\le
\ell(x)$. Further, each identity of a catoid admits the empty list as
a decomposition (of degree $0$).

Notions of length and $\ell$-categories originate in the work of
Mitchell~\cite{Mitchell72}. They have been used by
Leroux~\cite{Leroux75} in the context of Möbius categories.

\begin{lem}\label{lem:catoid-pump}
  Let $C$ be a catoid. Then for all $x,y\in C$, $Y\subseteq C$ and
  $n\in \mathbb{N}$,
  \begin{enumerate}
  \item $x\in x\odot y$ implies $x\in x\odot y^n$ and $x\in y \odot
    x$ implies $x\in y^n\odot x$,
  \item $x\in x\odot Y$ implies
  $x\in x\odot Y^n$ and $x\in Y\odot x$ implies
  $x\in Y^n\odot x$.
 \end{enumerate}
\end{lem}
\begin{proof}
  Item (1) follows from a simple induction on $n$; (2) is
  immediate from (1).
\end{proof}

\begin{lem}\label{lem:lcat-prop1}
  Let $C$ be an $\ell$-catoid. Then
  \begin{enumerate}
  \item  each identity in $C_0$ is indecomposable,
  \item $x\in x\odot y$ implies $y = t(x)$, and $x\in y\odot x$
    implies $y=s(x)$ for all $x,y\in C$,
    \item $x\in x \odot x$ implies $x\in C_0$.
  \end{enumerate}
\end{lem}
\begin{proof}
  For (1), suppose $x\in C_0$ is $n$-decomposable, hence in particular
  $x\in y \odot z$ for some $y,z\in C_1$. Then
  $x= s(x) = s(y\odot z)=s(y)$ and therefore
  $x\in x \odot (y \odot z)$, using the left identity axiom of
  catoids. Thus $x \in x \odot (y \odot z)^n$ for all
  $n\in\mathbb{N}$ by \autoref{lem:catoid-pump}, which contradicts
  $\ell(x)<\infty$. Every identity is therefore indecomposable.

  For (2), suppose $x \in x \odot y$. Then $x\in x\odot y^n$ for all
  $n\in \mathbb{N}$ by \autoref{lem:catoid-pump}, and
  $\ell(x)=\infty$ unless $y=t(x)$. The second property follows by
  opposition.

  Finally, (3) is immediate from (2).
\end{proof}

\begin{lem}\label{lem:lcat-prop2}
  A catoid $C$ is an $\ell$-catoid if
  \begin{enumerate}
  \item it is finitely $2$-decomposable,
  \item each identity in $C_0$ is indecomposable,
  \item $x\in x\odot y$ implies $y = t(x)$ for all $x,y\in C$.
  \end{enumerate}
\end{lem}
\begin{proof}
  Let $x\in C$, and suppose that the number of $2$-decompositions of
  $x$ is $k$. We claim that $x$ has length at most $k+1$. Indeed, if
  it is $(k+2)$-decomposable: $x=y_0\odot\cdots\odot y_{k+1}$, then we
  can obtain $k+1$ $2$-decompositions
  $y_0\odot(y_1\odot\cdots\odot y_{k+1}),\ldots, (y_0\odot\cdots\odot
  y_k)\odot y_{k+1}$.  Since there are only $k$ $2$-decompositions,
  two of these are the same. This means in particular that
  $y_0\odot\cdots\odot y_i \in(y_0\odot\cdots\odot
  y_i)\odot(y_{i+1}\odot\cdots\odot y_j)$ for some $i<j$.  But then
  $y_{i+1}\odot\cdots\odot y_j = t(y_i)$ by the the third assumption,
  which is impossible by the second assumption.
\end{proof}

\begin{lem}\label{lem:two-n-dec}
  In every catoid with indecomposable identities, each element is finitely
  $2$-decomposable if and only if each element is finitely
  $n$-decomposable for each $n\in\mathbb{N}$.
\end{lem}
\begin{proof}
  Consider a catoid $C$ in which each element is finitely
  $2$-decomposable. All elements in $C_0$ have a unique
  $0$-decomposition and no $n$-decompositions for $n>0$ by assumption.
  For elements in $C_1$ we proceed by induction on $n$. These elements
  have obviously no $0$-decomposition and unique $1$-decompositions
  (and $2$-decompositions by assumption). Hence suppose each element
  in $C_1$ is finitely $i$-decomposable for each $i \le n$. Each
  non-trivial $n+1$-decomposition of a given element in $C_1$ can
  obviously be seen in finitely many ways as a $2$-decomposition of
  two elements which both are finitely $i$-decomposable for all $i\le
  n$ by the induction hypothesis. This makes this element finitely
  $n+1$-decomposable. The converse implication is trivial.
\end{proof}

\autoref{lem:lcat-prop1} and \autoref{lem:lcat-prop2} can be combined as
follows, adapting a proposition of~\cite{ContentLL80} to catoids.

\begin{prop}\label{prop:moebius-eq}
  A catoid is Möbius if and only if
  \begin{enumerate}
  \item it is finitely $2$-decomposable,
  \item each identity is indecomposable,
  \item $x\in x\odot y$ implies $y=t(x)$.
  \end{enumerate}
\end{prop}
\begin{proof}
  If the catoid is Möbius, then (1) holds a fortiori and the other
  properties have been shown in
  \autoref{lem:lcat-prop1}. Conversely, properties (1), (2) and (3)
  imply that it is an $\ell$-catoid by
  \autoref{lem:lcat-prop2}. Property (1) and (2) imply that every
  element is finitely $n$-decomposable by \autoref{lem:two-n-dec}.
  So the catoid is Möbius.
\end{proof}

The following fact is the key to applications in the following
sections.

\begin{prop}
  A Möbius catoid is a finitely $2$-decomposable $\ell$-catoid.
\end{prop}
\begin{proof}
  Every Möbius catoid is obviously an $\ell$-catoid.  It is also
  finitely decomposable and therefore finitely $2$-decomposable. Every
  $\ell$-catoid satisfies properties (2) and (3) of
  \autoref{prop:moebius-eq} while property (1) from this
  proposition is assumed. Hence the catoid is is Möbius.
\end{proof}

The following properties of length are easy to check.
\begin{lem}\label{lem:length-prop}
  In every $\ell$-catoid $C$,
  \begin{enumerate}
  \item $\ell(x)=0$ if and only if $x\in C_0$,
  \item $\ell(x)\le 1$ for every indecomposable $x\in C$,
  \item $\ell(x)+\ell(y)\le \ell(z)$ for all $x,y\in C$ and
    $z\in x\odot y$.
 \end{enumerate}
\end{lem}

Adapting a definition by Mitchell, we say that an $\ell$-catoid
satisfies the \emph{saturated chain condition} if
$\ell(z)=\ell(x)+\ell(y)$ holds for all $x,y,z\in C$ such that
$z\in x\odot y$.

\begin{exa}\label{ex:moebius-catoids}~
 We now consider catoids and categories that have or lack the
 properties just discussed.
 \begin{enumerate}
 \item The free monoid and the shuffle monoid on the set $A$ from
   Examples~\ref{ex:catoids}(1) and (2) are Möbius and satisfy the
   saturated chain condition.
 \item The interval category $(I_P,\odot,s,t)$ on the poset $P$ from
   \autoref{ex:catoids}(3)is Möbius if each interval in $I_P$ is
   finitely $2$-decomposable~\cite{ContentLL80}. It need not satisfy
   the saturated chain condition. The Hasse diagram of the following
   finite poset shows that
   $\ell([a,b]) + \ell([b,c]) = 2 < 3 = \ell([a,c])$.
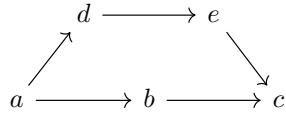
  \begin{equation*}
    \begin{tikzcd}[column sep = .4cm]
      & d\ar[rr] && e\ar[dr] &\\
      a\ar[ur]\ar[rr] && b\ar[rr] && c
    \end{tikzcd}
  \end{equation*}
  A poset $P$ is called \emph{locally finite} if every interval in
  $I_P$ is finitely $2$-decomposable ~\cite{Rota64}.
\item Pair groupoids (\autoref{ex:catoids}(4)) are generally not
  $\ell$-categories, even if the underlying set is finite. On the set
  $\{a,b,c\}$, we have
  $\ell((a,b))+\ell((b,c)) = 2 > 1 = \ell((a,b)\odot (b,c))$, where
  $\odot$ is the composition in the underlying pair groupoid.
\item The path category $C(G)$ on a graph $G$ from
  \autoref{ex:catoids}(5) forms an $\ell$-category that satisfies
  the saturated chain condition~\cite{Mitchell72}.  It need not be
  Möbius: the free category on the graph with vertices $a$, $b$, $c$,
  infinitely many edges from $a$ to $b$ and only single edge from $b$
  to $c$ is not finitely $2$-decomposable:
  
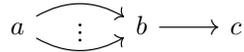
\begin{equation*}
    \begin{tikzcd}[column sep = .4cm]
      a\ar[rr, bend right]\ar[rr,bend left] &\raisebox{-2pt}{\vdots} & b\ar[rr] && c
    \end{tikzcd}
  \end{equation*}
  Yet $C(G)$ is Möbius if $G$ is finite.
\item The category $P(T,A)$ of guarded strings is Möbius and satisfies
  the saturated chain condition: $\ell(t)=0$ for $t\in T$,
  $\ell(a)=1$ for $a\in A$.
\end{enumerate}
\end{exa}

%%%%%%%%%%%%%%%%%%%%%%%%%%%%%%%%%%%%%%%%%%%%%%%%%%%%%%%

\section{Convolution semirings}\label{s:conv-algs}

Before turning to convolution algebras on catoids, we briefly list the
value algebras used in this construction: semirings, additively
idempotent semirings, Conway semirings, Kleene algebras and quantales,
with main emphasis on Kleene algebras.

A \emph{semiring} $(S,\cdot,+,0,1)$ consists of a monoid
$(S,\cdot, 1)$ and a commutative monoid $(S,+,0)$, such that
multiplication distributes over addition from the left and right and
$0$ is a left and right zero of multiplication. A \emph{dioid} is an
additively idempotent semiring: $\alpha+\alpha=\alpha$ holds for every
$\alpha\in S$.

In every dioid $S$, $(S,+,0)$ forms a join-semilattice with lattice
order $\le $ defined by $\alpha\le \beta\iff \alpha + \beta =\beta$ and least
element $0$. Multiplication preserves $\le$ in both arguments.

A \emph{Conway semiring}~\cite{Conway71,BloomE93} is a semiring $S$
with star $(-)^\ast:S\to S$ such that, for all
$\alpha,\beta\in S$,
\begin{equation*}
  (\alpha+\beta)^\ast = (\alpha^\ast\cdot \beta)^\ast\cdot \alpha^\ast\qquad \text{ and } \qquad 1+
  \alpha\cdot (\beta\cdot \alpha)^\ast \cdot \beta = (\alpha\cdot \beta)^\ast.
\end{equation*}
The second identity can be replaced by
$(\alpha\cdot \beta)^\ast \cdot\alpha = \alpha\cdot (\beta\cdot
\alpha)^\ast$, $1+\alpha\cdot \alpha^\ast = \alpha^\ast$ and
$1+\alpha^\ast \cdot \alpha = \alpha^\ast$, which simplifies
proofs in \autoref{s:conway-semirings}.

A \emph{Kleene algebra}~\cite{Kozen94} is a dioid $K$ with an
operation $(-)^\ast :K\to K$ such that
\begin{equation*}
  1+\alpha\cdot \alpha^\ast \le \alpha^\ast,\qquad \gamma +
  \alpha\cdot \beta\le \beta \implies
  \alpha^\ast \cdot \gamma \le \beta,\qquad \gamma + \beta\cdot \alpha
  \le \beta \implies \gamma \cdot
  \alpha^\ast \le \beta.
\end{equation*}

The first axiom is referred to as the unfold axiom, the others as
induction axioms. The opposite unfold axiom
$1+\alpha^\ast \cdot\alpha \le \alpha^\ast$ is derivable and so are
the identities $1+\alpha\cdot \alpha^\ast = \alpha^\ast$ and
$1+\alpha^\ast \cdot \alpha = \alpha^\ast$. Opposition means that one
formula is obtained from another by swapping the arguments in
multiplications. The class of Kleene algebras is closed under
opposition, and so are the other classes introduced in this section.

The induction axioms can be replaced by
\begin{equation*}
  \alpha \cdot \beta \le \beta \implies \alpha^\ast \cdot \beta\le \beta\qquad\text{ and
  }\qquad \beta\cdot \alpha \le
  \beta\implies \beta \cdot \alpha^\ast \le \beta,
\end{equation*}
which simplifies proofs in \autoref{s:convolution-kas}.

Standard identities for regular expressions, such as
$1\le \alpha^\ast$, $\alpha\cdot\alpha^\ast \le x^\ast$,
$\alpha^\ast \cdot\alpha \le \alpha^\ast$, $\alpha^i \le \alpha^\ast$
for all $i\in \mathbb{N}$ such that $\alpha^0=1$ and
$\alpha^{i+1} = \alpha\cdot\alpha^i$,
$\alpha^\ast \cdot\alpha^\ast = \alpha^\ast$,
$\alpha^{\ast\ast} = \alpha^\ast$,
$\alpha\le \beta \implies \alpha^\ast \le \beta^\ast$,
$(\alpha\cdot\beta)^\ast \alpha = \alpha \cdot
(\beta\cdot\alpha)^\ast$,
$(\alpha+\beta)^\ast = \alpha^\ast (\beta\cdot\alpha^\ast)^\ast$,
$\gamma\cdot\alpha\le \beta\cdot\gamma \implies \gamma\cdot
\alpha^\ast \le \beta^\ast\cdot \gamma$ and
$\alpha\cdot\gamma\le \gamma\cdot\beta\implies
\alpha^\ast\cdot\gamma \le \gamma\cdot\beta^\ast $ can be used to
reason with Kleene algebras. 

\begin{exa}\label{ex:value-ka}
  We list some typical value semirings or Kleene algebras, others
  can be found in the literature~\cite{DrosteK09}.
\begin{enumerate}
\item Any (bounded) distributive lattice is a dioid with inf as
  multiplication, in particular the distributive lattice $2$ of
  booleans with min as inf and max as sup. It extends to a Kleene
  algebra where the star is the constant $1$ map and $1$ the greatest
  element of the lattice.  The booleans thus form a Kleene algebra.
\item The max-plus semiring is defined on $\mathbb{R}^{-\infty}$ with
  $\max$ as addition, $+$ as multiplication, $-\infty$ as additive
  identity and $0$ as multiplicative identity. This defines a
  dioid. For a Kleene algebra one needs to restrict to the
  non-positive real numbers with $-\infty$ adjoined. The star is then
  the constant $0$ map.
\item The min-plus semiring is defined on $\mathbb{R}^\infty$ with
  $\min$ as addition, $+$ as multiplication, $\infty$ as additive
  identity and $0$ as multiplicative identity. To obtain a Kleene
  algebra one needs to restrict to the non-negative real numbers with
  $\infty$ adjoined. The star is then the constant $0$ map. This
  algebra is isomorphic to the Kleene algebra on $[0,1]$ with $\max$
  as addition, multiplication as multiplication, $0$ and $1$ as
  additive and multiplicative unit, and the constant $1$ map as the
  Kleene star.
\end{enumerate}
\end{exa}

We also need quantales. A \emph{quantale} $(Q,\le,\cdot,1)$ consists
of a complete lattice $(Q,\le)$ and a monoid $(Q,\cdot,1)$ such that
multiplication $\cdot$ preserves all sups in both
arguments~\cite{Rosenthal90}. We write $\bigvee$ for sups, $\bigwedge$
for infs as well as $\bot$ for the least and $\top$ for the greatest
element of the lattice. In any quantale we define
\begin{equation*}
  \alpha^\ast = \bigvee_{i \ge 0} \alpha^i
\end{equation*}
for $\alpha^0=1$ and $\alpha^{i+1} = \alpha \cdot \alpha^i$.  Using
this star it is routine to show that every quantale is a Kleene
algebra.

For the construction of convolution semirings, we fix a finitely
$2$-decomposable catoid $C$ and a semiring or dioid $S$. We write
$[P]$ for the indicator map of any proposition $P$, which evaluates to
$1 \in S$ if $P$ holds and to $0\in S$ otherwise.

We equip the set $S^C$ of functions $C\to S$ with a convolution
semiring structure:
\begin{itemize}
\item multiplication is convolution
  $*:S^C\times S^C\to S^C$: for any $f,g:C\to S$,
\begin{equation*}
  (f*g)(x)=\sum_{y,z\in C} f(y)\cdot g(z)\cdot [x\in y\odot z],
\end{equation*}
\item addition is defined by pointwise extension:
  $(f+g)(x) = f(x)+g(x)$,
\item the unit of multiplication is the set indicator function for
  $C_0$: $\id_0:C\to S,x\mapsto [x\in C_0]$,
\item the unit of addition is the constant zero map $0:C\to
  S,x\mapsto 0$.
\end{itemize}

The following fact is then standard.

\begin{thm}\label{thm:semiring}
  The set $S^C$ forms a convolution semiring (or convolution dioid).
\end{thm}

\begin{rem}
  For $S=2$, all functions in $2^C\simeq\Pow C$ are set-indicator
  functions. Convolution then specialises to the complex product
  $f\ast g= \bigcup \{y \odot z\mid y\in f, z\in g\}$ and convolution
  algebras become powerset algebras. The constructions of powerset
  algebras require no restrictions on catoids -- neither for complex
  products, nor for Kleene stars.
\end{rem}

\begin{rem}
  A convolution quantale $Q^C$ can be constructed along similar lines
  for an arbitrary catoid $C$ and any quantale $Q$, with convolution
  \begin{equation*}
    (f\ast g)(x)=\bigvee_{y,z\in C}f(y)\cdot g(z)\cdot [x\in y\odot z]
  \end{equation*}
  using arbitrary sups~\cite{Rosenthal97}. Further,
  $\id_0(x)= \bigvee_{y\in C_0} \delta_y(x)$ with
  $\delta_y(x)=[x=y]$.
\end{rem}

\begin{exa}\label{ex:convolution}
  We list some convolution semirings for a value semiring $S$ based on
  the catoids in Examples~\ref{ex:catoids} and
  \ref{ex:moebius-catoids}. All convolution semirings in these examples
  become convolution dioids if $S$ is a dioid, and convolution
  quantales if $S$ is a quantale.
\begin{enumerate}
\item For the free monoid on $A^\ast$ from
  \autoref{ex:catoids}(1), $S^{A^\ast}$ forms a semiring. Dioids
  of languages are obtained with value semiring $S=2$. Elements of
  $S^{A^\ast}$ are known as formal power series in language
  theory~\cite{DrosteK09}; they have also been studied in the
  combinatorics on words~\cite{CartierF69}.

\item For the shuffle catoid on $A^\ast$ from
  \autoref{ex:catoids}(2), $S^{A^\ast}$ forms a commutative
  semiring if $S$ is commutative.  Commutative dioids of shuffle
  languages are obtained with value algebra $S=2$.

\item For the interval category $I_P$ on a poset $P$ from
  \autoref{ex:catoids}(3), $S^{I_P}$ forms a semiring if $I_P$ is
  finitely $2$-decomposable. Maps from $I_P$ into a semiring, ring or
  field are known as \emph{incidence algebras}~\cite{Rota64}. For
  intervals over $\mathbb{N}$ or $\mathbb{Z}$ and value algebra $S=2$,
  this construction specialises to an interval temporal
  logic~\cite{Moszkowski12} with convolution as the chop operator, and
  without a next step operator~\cite{DongolHS21} (standard interval
  temporal logic admits only finite sups). Duration calculi, which are
  also considered in program correctness~\cite{ZhouH04}, use a similar
  chop operator on intervals.
  
\item For the pair groupoid on $X\times X$ from
  \autoref{ex:catoids}(4), $S^{X\times X}$ forms the semiring of
  $S$-valued or fuzzy binary relations~\cite{Goguen67} if $X$ is
  finite. The standard dioid of binary relations on $X$ is
  obtained with $S=2$. Binary relations correspond to graphs and boolean-valued
  matrices. Weighted binary or $S$-valued relations thus correspond to weighted graphs and
  to matrices with convolution as matrix multiplication.

\item For the path category $C(G)$ on graph $G$ from
  \autoref{ex:catoids}(5), $S^{C(G)}$ forms a semiring if $G$ is
  finite. An analogous result holds for convolution semirings on the
  path category $P(T,A)$ of guarded strings.
\end{enumerate}
\end{exa}

Convolution algebras similar to those in this section are widely
studied in mathematics, but semirings are usually replaced by rings
and catoids by monoids, groups, categories or groupoids. The resulting
algebras are known as monoid algebras, group algebras, category
algebras or groupoid algebras.

%%%%%%%%%%%%%%%%%%%%%%%%%%%%%%%%%%%%%%%%%%%%%%%%%%%%%%%%%

\section{Convolution Kleene algebras}\label{s:convolution-kas}

We now extend a previous recursive construction of the Kleene star for
convolution algebras with relational monoid objects with one single
identity~\cite{CranchDS21} to general catoids with multiple
identities. We also replace a previous ad hoc grading that allows
induction on certain catoids with one single identity by the more
structural and liberal Möbius condition. Our construction of the star
generalises that of Kuich and Salomaa~\cite{KuichS86} beyond formal
power series on free monoids.  Their value semirings are also quite
different from Kleene algebras.  Ésik and Kuich have used Kuich and
Salomaa's construction to show that formal power series on free
monoids with inductive $\ast$-semirings as value algebras form
inductive $\ast$-semirings~\cite{EsikK04}. Inductive $\ast$-semirings
are more similar to Kleene algebras: they are ordered, but not
necessarily additively idempotent semirings in which the first star
unfold axiom and the first star induction axiom of Kleene algebras
hold, but not their duals.

We fix a Möbius catoid $C$ and a Kleene algebra $K$. For all
$f:C\to K$, $x\in C_1$ and $e\in C_0$ we define
\begin{equation}
\label{eq:star}
f^\ast (e)= f(e)^\ast, \qquad f^\ast (x) = f(s(x))^\ast \cdot \sum_{y,z\in
  C} f(y)\cdot f^\ast (z) \cdot [x\in y \odot z, y\neq s(x)].
\end{equation}

It is easy to show by a simple induction on $\ell(x)$ that, for $x\in C_1$,
\begin{equation}
  \label{eq:star-intuitive}
  f^\ast(x) = \sum_{1\le i \le \ell(x)}\sum_{x_1,\dots,x_i\in C_1}
  \left(\prod_{1\le j\le i} f(s(x_j))^\ast\cdot
f(x_j)\right) \cdot f(t_i(x_i))^\ast [x\in x_1 \odot x_2 \odot \dots
\odot x_i].
\end{equation}
This more symmetric formula suggests that the second identity in
\eqref{eq:star} has a dual.

\begin{lem}\label{lem:starswap}
  For all $f:C\to K$ and $x\in C_1$,
  \begin{equation*}
    f^\ast (x) = \left(\sum_{y,z\in C} f^\ast (y)\cdot f(z)
    \cdot [x\in y \odot z, z\neq t(x)]\right) \cdot f(t(x))^\ast.
\end{equation*}
\end{lem}
\begin{proof}
  We proceed by induction on $\ell(x)$, dropping all multiplication
  symbols in $C$ and $K$.
\begin{align*}
  &f^\ast (x)\\
  &= f(s(x))^\ast \sum_{u,y\in C} f(u) f^\ast (y) [x\in u y, u\neq s(x)]\\
&= f(s(x))^\ast  \sum_{u,y \in C} f(u)\left(\sum_{v,w\in C}f^\ast
                                                                            (v)f(w)[y\in vw, w\neq t(y)]\right) f(t(y))^\ast [x\in uy,u \neq s(x)]\\
&=  \sum_{y,u,v,w \in C} f(s(x))^\ast f(u) f^\ast (v) f(w)
                                                                                                                                                       f(t(x))^\ast[y\in vw, w\neq t(x),x\in uy, u\neq s(x)] \\
&=  \sum_{z,u,v,w \in C} f(s(x))^\ast f(u) f^\ast (v) f(w) f(t(x))^\ast[z\in uv, w\neq t(x),x\in zw, u\neq s(x)] \\
&=  \left(\sum_{z,w\in C} \left(f(s(z))^\ast \sum_{u,v\in
                                                               C}f(u)f^\ast
                                                               (v) [z\in
             uv,
             u\neq
             s(z)]
                                                               \right)
                                                               f(w) [x\in z w, w\neq t(x)]\right)
                                                               f(t(x))^\ast \\
           &= \left(\sum_{z,w\in C} f^\ast(z)f(w)[x\in zw,  w\neq t(x)]\right)f(t(x))^\ast.
\end{align*}
The first and last step use the definition of $f^\ast$.  The second
applies the induction hypothesis with $\ell (x)>\ell(y)$. The third
uses distributivity laws and $t(x)=t(y)$, which holds because
$x\in u\odot y$ using \autoref{lem:catoid-props}.8.  The fourth step
uses associativity in $C$. The fifth applies distributivity laws and
$s(x)=s(z)$, using $x\in z\odot w$ and again
\autoref{lem:catoid-props}.8.
\end{proof}

We are now prepared for our main result.

\begin{thm}\label{thm:main-theorem}
  The set $K^C$ forms a convolution Kleene algebra with star
  given by \eqref{eq:star}.
\end{thm}

\begin{proof}
  As $K^C$ forms a dioid (\autoref{thm:semiring}), it remains to check
  the Kleene algebra axioms
\begin{equation*}
\id_0+ f * f^\ast \leq f^\ast, \qquad f*g \leq g \implies f^\ast *
g\leq g, \qquad g*f\leq g \implies g*f^\ast \leq g
\end{equation*}
on the convolution algebra. Again we drop all multiplication symbols
in $C$, $K$ and $K^C$.

For the star unfold law $\id_0(x)+ (f f^\ast)(x) \leq f^\ast(x)$ we
proceed by case analysis on $x\in C$. We abbreviate
\begin{equation*}
  \Sigma_\ast(x) = \sum_{y,z\in C} f(y)\cdot f^\ast (z) \cdot [x\in y
  \odot z, y\neq s(x)].
\end{equation*}
If $x\in C_0$, then
\begin{equation*}
  \id_0(x)+ (ff^\ast)(x)=1 + f(x)f^\ast(x)=f^\ast(x),
\end{equation*}
using the star unfold law in $K$, because $x$ is indecomposable and
hence $f(x)f^\ast(x)$ is the only summand in the
convolution. Otherwise, if $x\in C_1$, then
\begin{align*}
  \id_0(x)+ (ff^\ast)(x)
  &= \sum_{y,z\in C} f(y)f^\ast(z)[x\in yz] \\
  &= f(s(x)) f^\ast (x)+ \Sigma_\ast(x)\\
  &=  f(s(x))f(s(x))^\ast \Sigma_\ast(x)+\Sigma_\ast(x)\\
  &= \left(f(s(x))f(s(x))^\ast +\id_0(s(x))\right)\Sigma_\ast(x)\\
  & = f(s(x))^\ast\Sigma_\ast(x)\\
  &= f^\ast(x).
\end{align*}

In the first step $id_0(x)=0$ because $x\notin C_0$. The second step
rearranges the summation. The third unfolds the definition of the star
in $K^C$ in the first summand. The remaining steps use laws from $K$.

For the first star induction law we assume that $fg\leq g$, that is
$(fg)(x)\leq g(x)$ for all $x\in C$, and we show that
$f^\ast g\leq g$, that is $(f^\ast g)(x)\leq g(x)$ for all $x\in C$ by
induction on $\ell(x)$.

If $\ell(x)=0$ and hence $x\in C_0$, we have
$(fg)(x)=f(x)g(x)\le g(x)$ because identities are indecomposable. Thus
$(f^\ast g)(x)=f^\ast(x) g(x)\le g(x)$, using again indecomposability
of identities and the (simplified) star induction law in $K$.

For $\ell(x)\ge 1$, and thus $x\in C_1$, suppose
$(f^\ast g)(y)\leq g(y)$ for all $y$ such that
$\ell(y)<\ell(x)$. Moreover, the assumption of star induction implies
that for all $y,z$ such that $x\in yz$, we have $f(y)g(z)\leq g(x)$,
from which $f(s(x))^\ast g(x)=f^\ast(s(x)) g(x)\leq g(x)$ follows
using star induction in $K$. We then calculate
\begin{align*}
  &(f^\ast g)(x)\\
 % &=f^\ast (s(x)) g(x)+\sum_{y,z\in C}f^\ast (y) g(z)[x\in yz, y\neq s(x)]\\
  &=f^\ast (s(x))g(x)+\sum_{y,z\in C}\left(f(s(y))^\ast \sum_{u,v\in C} f(u) f^\ast (v) [y\in uv, u\neq s(y)]\right)g(z)[x\in yz, y\neq s(x)] \\
  % &=f^\ast (s(x)) g(x)+\sum_{y,z\in C}\left(f(s(x))^\ast \sum_{u,v\in C} f(u)f^\ast (v) [y\in uv, u\neq s(x)]\right)g(z)[x\in yz, y\neq s(x)]\\
  &=f^\ast (s(x))\left( g(x)+\sum_{y,z,u,v\in C}
    f(u)f^\ast
    (v)
    g(z)
    [y\in
    uv, u\neq s(x), x\in yz, y\neq s(x)]\right) \\
  &=f^\ast (s(x))\left( g(x)+\sum_{y,z,u,v\in C} f(u)(f^\ast (v)g(z))
    [y\in vz, u\neq s(x), x\in uy,y\neq s(x)]\right)\\
  &=f^\ast (s(x)) \left( g(x)+\sum_{u,y\in C} f(u )\left(\sum_{v,z\in C}
    (f^\ast (v)g(z)) [y\in vz]\right)[u\neq s(x), x\in uy,y\neq
    s(x)]\right)\\
  &=f^\ast (s(x)) \left( g(x)+\sum_{u,y\in C} f(u )(f^\ast g)(y)[u\neq s(x), x\in uy,y\neq s(x)]\right)\\
  &\leq f^\ast (s(x)) \left( g(x)+\sum_{u,y\in C} f(u )g(y)[x\in uy]\right) \\
 &= f^\ast(s(x))(g(x)+(fg)(x))\\
%  &\leq f^\ast (s(x))\left( g(x)+g(x)\right)\\
  &\leq f(s(x))^\ast g(x) \\
  &\leq  g(x).
\end{align*}
The first step rearranges the summation and unfolds the definition of
$f^\ast$. The second applies $s(y)=s(x)$, using $x\in y \odot z$ and
\autoref{lem:catoid-props}.8, and distributivity laws. The third
uses associativity in $C$, the fourth distributivity and the fifth the
definition of convolution. The sixth step applies the induction
hypothesis, using that $\ell(u)+\ell(y)\le \ell(x)$, it also relaxes
some summation constraints. The eight uses again the definition of
convolution, the ninth the assumption $(fg)(x)\le g(x)$ and
idempotency of addition, and the last star induction in $K$, as
outlined before this calculation.

The proof of the second star induction law follows by opposition from
that of the first, using \autoref{lem:starswap} in the
induction step to unfold $f^\ast$.
\end{proof}

\begin{exa}\label{ex:conv-kas}
  The catoids in Examples~\ref{ex:catoids} and
  \ref{ex:moebius-catoids} yield examples where the recursive
  definition \eqref{eq:star} of the Kleene star in convolution Kleene
  algebras can be used and where this is not the case, indicating the
  potential and limitations of this construction. 
\begin{enumerate}
\item For the free monoid on $A^\ast$ from
  \autoref{ex:catoids}(1), $K^{A^\ast}$ forms the Kleene algebra
  of formal power series on $A^\ast$, a well known result. Language
  Kleene algebras~(\cite{Kozen94}) are obtained for $K=2$.

\item For the shuffle catoid on $A^\ast$ from
  \autoref{ex:catoids}(2), $K^{A^\ast}$ forms a commutative Kleene
  algebra if $K$ is commutative, another classical result. In this and
  the previous example, the empty word is the only unit and the
  definition of the star in \eqref{eq:star} is recursive with respect
  to the length of words. Möbius catoids comprise this case, but
  their full expressivity is not needed. Commutative Kleene algebras
  of shuffle languages are obtained for $K=2$.

\item The finitely $2$-decomposable interval categories $I_P$ from
  \autoref{ex:catoids}(3) are Möbius categories with many units. Hence
  $K^{I_P}$ forms a Kleene algebra -- the \emph{incidence Kleene
    algebra} on $P$. The special case $K=2$ yields interval temporal
  logics (over finite intervals) with chop-star
  operator~\cite{Moszkowski12} given by $f^\ast$, but without a
  next-step operator.  The lack of a suitable Kleene star on the
  convolution algebra prevented a Kleene algebraic weighted treatment
  of this logic using finite sups so far -- except for $K=2$, where
  the Möbius condition is not needed~\cite{DongolHS21}. Now the Kleene
  star $f^\ast$ \autoref{thm:main-theorem} supplies a suitable chop-star
  operator, which is also available in duration
  calculi~\cite{ZhouH04} and other interval logics. 

\item Convolution dioids on pair groupoids (\autoref{ex:catoids}(4))need
  not generalise to matrix Kleene algebras or Kleene algebras of
  weighted relations using the recursive star à la Kuich and
  Salomaa. Recall that even finite pair groupoids generally have no non-trivial
  length. The star on matrix Kleene algebras and Kleene algebra of
  weighted relations is therefore defined by other
  means~\cite{Conway71}:
  \begin{equation*}
    \begin{pmatrix}
      A & B\\
      C & D
    \end{pmatrix}^\ast
    =
    \begin{pmatrix}
      (A+ BD^\ast C)^\ast & A^\ast(B(D+CA^\ast B)^\ast\\
      D^\ast C(A+ BD^\ast C)^\ast & (D+CA^\ast B)^\ast
    \end{pmatrix},
  \end{equation*}
  where the matrix under the star is partitioned into submatrices $A$,
  $B$, $C$ and $D$ such that $A$ and $D$ are square, and to which the
  definition of the star is applied recursively.
\item The path category $C(G)$ on a finite graph $G$ from
  \autoref{ex:catoids}(5) forms again a Möbius category with many
  units; $K^{C(G)}$ forms a Kleene algebra. Likewise for the category
  $P(T,A)$ of
  guarded strings.
  \end{enumerate}
\end{exa}

\begin{rem}
  All convolution Kleene algebras in Example~\ref{ex:conv-kas} have
  computational content. Languages with language and shuffle product
  are relevant to the interleaving semantics of concurrent
  systems. Formal power series with a sequential and a shuffle
  convolution provide quantitative variants, which are studied further
  in Section~\ref{s:concurrent-convolution-ka}. Incidence Kleene
  algebras form the basis of quantitative interval temporal logics,
  duration calculi and similar temporal logics that use a binary chop
  operator. Binary relations form a standard semantics of sequential
  programs; weighted relations have important applications in fuzzy
  mathematics, probabilistic program semantics and quantitative
  program correctness. Our results explain why the star of Kuich and
  Salomaa cannot be used for such applications. Finally, path
  categories provide notions of execution traces of programs, notions
  of rewriting sequences in abstract rewriting systems and the data
  underlying graph algorithms. These are explored in the remaining
  sections of this article.
\end{rem}

The following lemma specialises the second equation in \eqref{eq:star}
to powers, giving a formula that captures the precise decomposition of the two
stars in \eqref{eq:star} for $f^n$ instead of $f^\ast$.  This
translates a similar statement from language theory~\cite{KuichS86} to
catoids and dioids, and will be used in the treatment of $\ast$-continuous Kleene algebras in~\autoref{A:star}.

\begin{lem}
  \label{lem:conv-star-prop}
  Let $C$ be a finitely $2$-decomposable catoid and $S$ a dioid. 
  Then for all $f:C\to S$,
  $x\in C_1$ and $n\ge 1$,
  \begin{equation}
    \label{eq:power-exact}
    f^n(x) = \sum_{y,z\in C} \sum_{i=0}^{n-1} f^i(s(x))
    f(y)f^{n-1-i}(z)[x\in y\odot z,y\neq s(x)].
  \end{equation}
\end{lem}
\begin{proof}
  By induction on $n$. We abbreviate $I(x,y,z)=[x\in y\odot z,y\neq
  s(x)]$.

  For $n=1$,
  \begin{equation*}
    \sum_{y,z\in C} \sum_{i=0}^{1-1} f^i(s(x))
    f(y)f^{1-1-i}(z) I(x,y,z)
   = \sum_{y,z\in C}f^0(s(x))f(y)f^0(z)I(x,y,z) = f^1(x).
 \end{equation*}
 For $n+1$,
 \begin{align*}
   f^{n+1}(x)
   &=\sum_{y,z\in C} f(y)f^n(z)[x\in y\odot z]\\
   &=\sum_{y,z\in C} f(y)f^n(z)I(x,y,z) +
     f(s(x))f^n(x)\\
   &=\sum_{y,z\in C} f(y)f^n(z)I(x,y,z) + f(s(x)) \sum_{y,z\in C} \sum_{i=0}^{n-1} f^i(s(x))
     f(y)f^{n-1-i}(z)I(x,y,z)\\
   &=\sum_{y,z\in C} f(y)f^n(z)I(x,y,z) + \sum_{y,z\in C} \sum_{i=0}^{n-1} f^{i+1}(s(x))
     f(y)f^{n-1-i}(z)I(x,y,z)\\
   &=\sum_{y,z\in C} f(y)f^n(z)I(x,y,z) + \sum_{y,z\in C} \sum_{i=1}^{n} f^{i}(s(x))
     f(y)f^{n-i}(z)I(x,y,z)\\
   &=\sum_{y,z\in C} \sum_{i=0}^{n} f^i(s(x))
     f(y)f^{n-i}(z)I(x,y,z).\qedhere
 \end{align*}
\end{proof}

%%%%%%%%%%%%%%%%%%%%%%%%%%%%%%%%%%%%%%%%%%%%%%%%%%%%%%%%%%%

\section{Special functions in convolution Kleene algebras}
\label{s:path-algorithms}

Applications such as path algorithms require convolution algebras
where all identities in a path category $P(C)$ have value $1$ in the
Kleene algebra. Another interesting set of functions are indicator
functions for sets of identities of catoids. We now study these
special functions on convolution Kleene algebras.

Let $C$ be a catoid, $S$ a dioid and $K$ a Kleene algebra. Consider
the sets $S[C] \subseteq S^C$ and $K[C] \subseteq K^C$ of functions
that contain the zero map $0$ and in which all $f\neq 0$ satisfy
$f(e)=1$ for all $e\in C_0$.

For every Möbius catoid $C$ and $f\in K[C]$, the star in
\eqref{eq:star} specialises, for all $e\in C_0$ and $x\in C_1$, to
\begin{equation}
   \label{eq:star-path}
  f^\ast (e)= 1, \qquad f^\ast (x) = \sum_{y,z\in
    C} f(y)\cdot f^\ast (z) \cdot [x\in y \odot z, y\neq s(x)].
\end{equation}

Under the obvious conditions on $C$, the sets $S[C]$ and $K[C]$ form a
convolution sub-dioid and a sub-Kleene algebra of $S^C$ and $K^C$,
respectively.

\begin{prop}
  \label{prop:conv-dioid-special}
   \label{prop:conv-ka-special}
  Let $C$ be a catoid, $S$ a dioid and $K$ a Kleene algebra. Then
  \begin{enumerate}
  \item $S[C]$ forms a
    dioid if $C$ is finitely $2$-decomposable,
    \item $K[C]$ forms a Kleene algebra if $C$ is Möbius.
  \end{enumerate}
 \end{prop}
\begin{proof}
  For (1),  we first show that $S[C]$ is closed under the semiring
  operations. Obviously, $0$ and $\id_0$ are in $S[C]$ and if
  $f,g\in S[C]$, then so is $f+g$, because addition is
  idempotent in every dioid.  For closure under $\ast$, $f\ast g = 0$ if at least one
  of $f$ and $g$ is $0$. Otherwise, if $f\neq 0\neq g$ and $x\in C_0$,
  then
\begin{align*}
  (f \ast g)(x)
  = \sum_{y,z} f(y)\cdot g(z)\cdot [x\in y\odot z]
  = f(x)\cdot g(x) = 1.
\end{align*}
Thus $S[C]$ is closed under convolution as
well. \autoref{thm:semiring} gives us a dioid structure on $S^C$,
hence the dioid laws hold in particular in $S[C]$.

For (2),  it remains to show that $f^\ast \in K[C]$ whenever $f$
is. It is immediate from the definition of the star that
$0^\ast = \id_0\in K[C]$.  Otherwise for all $f\neq 0$ and for $x\in C_0$,
$f^\ast(x) =f(x)^\ast = 1^\ast =1$. \autoref{thm:main-theorem}
then implies that $K[C]$ forms a Kleene algebra.
\end{proof}

Further, under the same conditions as above, \eqref{eq:star-intuitive}
simplifies to
\begin{equation}
  \label{eq:star-intuitive-path}
  f^\ast(x) = \sum_{1\le i \le \ell(x)}\sum_{x_1,\dots,x_i\in
    C_1}\left(\prod_{1\le j\le i}
f(x_j)  \right)\cdot [x\in x_1 x_2 \dots x_i].
\end{equation}
It thus computes the optimal value among the non-identity
decompositions of $x$.

\begin{exa}
  For $C=P(G)$, the function $f^\ast(\pi)$ simply computes the value,
  weight or cost of a path $\pi$ in the graph $G$ as the product of
  the weights of its edges.
\end{exa}

\begin{rem}
  The general path problem, for a finite directed graph $G$, asks for
  computing the optimal weight on any set of paths between two given
  vertices, starting from weights on the edges in $G$. Typical
  instances are transitive closure or shortest path algorithms. The
  original algebraic approach by Aho, Hopcroft and Ullman uses
  implicitly a convolution algebra on $P(G)$ to compute weights on
  homsets $P(G)(v,v')$ for each pair of vertices $v$, $v'$ in
  $G$~\cite[Section 5.6]{AhoHU75}. Weights are taken in closed
  semirings, essentially quantales in which only countable sups are
  assumed (for path problems this restriction seems
  insignificant). The approach has been generalised to non-idempotent
  closed semirings~\cite[Chapter V]{Mehlhorn84}; see
  also~\cite{Mohri02} for a survey of algorithmic examples. Infinite
  sups or sums are needed for extending weights from paths to homsets,
  which can be infinite in graphs with cycles. Kleene algebras or
  Conway semirings can be used for finite directed acyclic graphs,
  which are relevant, for instance, in causal inference, Bayesian
  networks or the provenance analysis of finite
  games~\cite{GraedelT20}.

  More generally, for a convolution quantale $Q^C$ on a catoid $C$, the
  general path problem amounts to computing
  \begin{equation*}
    f(e,e') = \bigvee_{x\in C(e,e')}f^\ast(x)
  \end{equation*}
  for $C(e,e') =\{x\in C\mid s(x)=e\land t(x)=e'\}$ and
  $e,e' \in C_0$. Alternatively, one can use $K[C]$ with $f^\ast$
  given by \eqref{eq:star-path} if $C$ is finite, as in the case of
  DAGs.

  The two classical solutions to the general path problem are based
  on a variant of Kleene's algorithm (for constructing regular
  expressions from automata) and on Conway's algorithm for
  constructing the star of a matrix outlined in
  \autoref{ex:conv-kas}(4). Both assemble the $f(e,e')$ into a
  matrix (as in \autoref{ex:conv-kas}(4)), identifying $e$ and
  $e'$ with vertices in $G$.  They construct this weight matrix
  starting from matrices that consider edge weights only, setting all
  other weights to zero. These bottom-up approaches make the
  construction of $f^\ast$, using \eqref{eq:star-path},
  unnecessary. Kleene's algorithm generally requires constructing the
  star in the weight algebra, but not in particular cases such as
  transitive closure or basic shortest path algorithms. Conway's
  algorithm requires only the computation of the matrix star.  The
  recursive star defined in \eqref{eq:star} or \eqref{eq:star-path}
  may be interesting for computing $f(e,e')$ with catoids beyond
  $P(G)$, but this remains beyond the scope of this article.
  \end{rem}

Another interesting subset of functions in $S^C$ and $K^C$ are the
indicator functions $\chi_A: x\mapsto [x\in A]$ for each
$A\subseteq C$ in a Möbius catoid $C$, and in particular the indicator
functions $\chi_P$ for $P\subseteq C_0$. We write
\begin{equation*}
  \chi=\{\chi_A\mid A\subseteq C\}\qquad\text{ and }\qquad \chi_0 = \{\chi_P\mid P\subseteq C_0\}.
\end{equation*}
Obviously, $\chi_0 = \{f\in \chi\mid f\le \id_0\}\subseteq \chi$,
$\chi_0\subseteq \{f\in C\mid f\le \id_0\}$ and $\chi_{\{x\}} =
\delta_x$ for each $x\in C$.

\begin{rem}
  Let $C$ be a catoid and $Q$ a quantale. Then any $f:C\to Q$
  satisfies $f=\bigvee_{x\in C} f(x)\cdot\delta_x$, sampling the values of
  $f$ point-wise. Thus in particular
  $\chi_A=\bigvee_{x\in A}\delta_x$ for all $A\subseteq C$ and
  $\id_0 = \bigvee_{x\in C_0} \delta_x$.
\end{rem}

\begin{prop}
  \label{prop:indicator-ka}
  Let $C$ be a Möbius catoid and $K$ a Kleene algebra.  Then $\chi$
  forms a sub-Kleene algebra of $K^C$ isomorphic to the powerset
  Kleene algebra $\Pow C$.
\end{prop}
\begin{proof}
  Given \autoref{thm:main-theorem}, we need to show that indicator
  functions form a sub-Kleene algebra and establish the isomorphism.
  First, for every dioid $S$, it is easy to check that
  $\chi\subseteq S^C$ contains $0$ and $\id_0$ and is closed under
  addition and convolution whenever $C$ is finitely
  $2$-decomposable. It is thus a sub-dioid of $S^C$ that is isomorphic
  to the powerset dioid $\Pow C$. Further, under the conditions of the
  proposition, formula \eqref{eq:star-intuitive} for $f^\ast$
  specialises to \eqref{eq:star-intuitive-path}, as
  $0^\ast = 1^\ast = 1$ in every Kleene algebra. Thus $f^\ast$ is an
  indicator function whenever $f$ is.

  Second, $f^\ast(x)=1$ if and only
  if $x\in x_1\dots x_i$ for some $i$ such that $f(x_j)=1$ for all
  $1\le j\le i$. Thus $x$ is in the set represented by $f^\ast$ if and
  only if $x\in x_1\dots x_i$ and, for $1\le j \le i$, every $x_j$ is
  in the set represented by $f$, which is the standard definition of
  the Kleene star in the powerset Kleene algebra $\Pow C$.
\end{proof}

Relating $\chi_0$ with $K^C$ requires a definition. A \emph{Kleene
  algebra with tests}~\cite{Kozen97} is a two-sorted structure formed
by a Kleene algebra and a boolean algebra (the test algebra) which is
embedded into the Kleene algebra such that $0$ is the least and $1$
the greatest element of the boolean algebra and binary sups and infs
of elements in the boolean algebra are sent to binary sums and
products of elements of the Kleene algebra.

\begin{thm}
 \label{thm:main-theorem-kat}
 Let $C$ be a Möbius catoid and $K$ a Kleene algebra. Then $K^C$ forms
 a Kleene algebra with tests with (atomic boolean) test algebra $\chi_0$.
\end{thm}
\begin{proof}
  In every Kleene algebra, the elements below $1$ form a sub-Kleene
  algebra, and the star of each such element equals $1$. Hence, by
  \autoref{prop:indicator-ka}, $\chi_0$ forms a sub-Kleene
  algebra of the sub-Kleene algebra $\chi \simeq\Pow C$ of $K^C$.
  Further, it is straightforward to check that $\chi_0$ forms an atomic
  boolean algebra in which binary sup in $\chi$ is addition, binary
  inf convolution, $\chi_\emptyset$ is the constant zero function $0$
  and $\chi_{C_0}=\id_0$. The boolean complement of $\chi_P$ is
  $\chi_{C_0-P}$, the atoms are the functions $\delta_x$ for all
  $x\in C_0$.
\end{proof}

This result generalises a previous construction of Kleene algebras
with tests on formal power series on guarded strings by
Sedlár~\cite[Lemma 1]{Sedlar24} along the lines of an approach to
formal powerseries on words forming algebras similar to Conway
semirings by Ésik and Kuich~\cite{KuichE01}. In this approach, finite
semirings are used as value algebras and functions are required to
have finite support. Note that a module-like approach is used, in
which functions can be multiplied with weights from the semiring.  We
ignore this scalar multiplication, but could add it easily.  Apart
from that, our approach is more general on the domain and
codomain of functions, as well as on the class of functions itself.

\begin{rem}\label{rem:test-trivial}
  In $K[C]$, $\chi_0=\{0,\id_0\}$; hence the test algebra of the
  Kleene algebra with tests $K[C]$ is trivial.
\end{rem}

\begin{exa}
  We reconsider the structures in \autoref{ex:conv-kas} in light of
  \autoref{thm:main-theorem-kat}. While the convolution Kleene algebra
  with tests $K^{A^\ast}$ on the free monoid or the shuffle catoid on
  $A^\ast$ have trivial test algebras as in \autoref{rem:test-trivial},
  the test algebras of the incidence Kleene algebras $K^{I_P}$ are
  formed by the indicator maps for the subsets of identity intervals
  in $I_{P}$. Likewise, in $K^{P(G)}$, the test algebras are formed by
  the indicator maps for the subsets of vertices of $G$.  Most
  interesting in programming application of Kleene algebras with tests
  are categories of guarded strings and pair groupoids. For the
  former, the situation is similar to $P(G)$; for the latter, the star
  cannot be defined via \eqref{eq:star}, see
  \autoref{ex:conv-kas}(4). Yet convolution dioids on finitely
  $2$-decomposable pair catoids have a natural test structure given by
  subsets of the identity relation $\id_0$.
\end{exa}

%%%%%%%%%%%%%%%%%%%%%%%%%%%%%%%%%%%%%%%%%%%%%%%%%%%%%%%%%%

\section{Modal convolution Kleene
  algebras}\label{s:modal-convolution-ka}

In this and the following two sections we present extensions of
\autoref{thm:main-theorem}.  Semirings, Kleene algebras and quantales
have been equipped with domain and codomain operators, inspired by
algebras of binary relations. The resulting modal semirings and Kleene
algebras allow defining predicate transformer algebras akin to
propositional dynamic logics and can be applied in program
verification~\cite{DesharnaisS11,GomesS16}.  For every local catoid
$C$ and quantale $Q$ with domain and codomain operations, $Q^C$ forms
such a quantale~\cite{FahrenbergJSZ23}. Here, we use
\autoref{thm:main-theorem} to specialise this convolution quantale
construction to convolution Kleene algebras with domain and codomain
operations, applicable to weighted program verification.

A \emph{modal semiring}~\cite{DesharnaisS11} is a dioid $S$ with
domain and codomain maps $\dom,\cod:S\to S$ that satisfy, for all
$\alpha,\beta\in S$,
\begin{itemize}
\item for the domain map:
\begin{gather*}
\alpha\le \dom(\alpha)\cdot \alpha,\qquad \dom(\alpha\cdot \dom(\beta)) = \dom(\alpha\cdot \beta),\qquad
\dom(\alpha)\le 1,\\
\dom(0)=0,\qquad \dom(\alpha+\beta) = \dom(\alpha)+\dom(\beta),
\end{gather*}
\item for the codomain map, by opposition, $\alpha\le \alpha\cdot
  \cod(\alpha)$, $\cod(\cod(\alpha)\cdot \beta) = \cod(\alpha\cdot
  \beta)$, $\cod(\alpha)\le 1$, 
$\cod(0)=0$ and $\cod(\alpha+\beta) = \cod(\alpha)+\cod(\beta)$, 
\item for both maps: $\cod(\dom(\alpha))=\dom(\alpha)$ and $\dom(\cod(\alpha))=\cod(\alpha)$.
\end{itemize}
Opposition means that the codomain axioms are obtained from the domain
ones by swapping the arguments in multiplications and exchanging
$\dom$ and $\cod$.

A \emph{modal Kleene algebra}~\cite{DesharnaisS11} is a modal semiring
that is also a Kleene algebra.

In every modal Kleene algebra, the set $K_0$ of fixpoints of $\dom$,
which equals the set of fixpoints of $\cod$, forms a subalgebra of
$K$, which is a distributive lattice bounded by $0$ and $1$ in which
$+$ is binary sup and $\cdot$ is binary inf. Similarly, we write $S_0$
in case of modal semirings. The lattices $S_0$ and $K_0$ can be
identified with algebras of propositions or tests on which programs
act via modal operators: for instance, for $\alpha\in S$ and
$p\in S_0$,
\begin{equation*}
  |\alpha\rangle p =\dom(\alpha\cdot p)\qquad\text{and}\qquad\langle \alpha|p = \cod(p\cdot \alpha)
\end{equation*}
are forward and backward diamond operators.

Domain and codomain operations have been extended to modal convolution
quantales $Q^C$ where the value quantale $Q$ is equipped with a domain
and codomain operation satisfying the same axioms as for modal
semirings~\cite{FahrenbergJSZ23}: for all $x\in C$ and $f:C\to Q$,
\begin{equation*}
  \Dom(f)(x)=\bigvee_{y\in C}\dom(f(y))\cdot
  \delta_{s(y)}(x)\qquad\text{ and }\qquad\Cod(f)(x)=\bigvee_{y\in
    C}\cod(f(y))\cdot \delta_{t(y)}(x).
\end{equation*}

\begin{exa}\label{ex:Domain}
  In the construction of convolution quantales, the
  source target structure of the catoid $C$ is not fully reflected:
  $\id_0(x)=[x\in C_0]$ conflates all elements in $C_0$ in $1\in Q$
  (and all other elements in $0$). Modal quantales capture the source
  and target structure in terms of the domain and codomain operations,
  as the following two examples show.
  \begin{enumerate}
  \item For $Q=2$, $\Dom(f)(x)$ indicates whether $x$ is an element
    of the set represented by $\Dom(f)$. This is the case if $x$ is the
    source of some element $y$ in the set represented by $f$. As
    $\dom(1)=1$ and $\dom(0)=0$ in any domain semiring, the set
    represented by $\Dom(f)$ is therefore the image of the set
    represented by $f$ under $s$. Likewise, the set represented by
    $\Cod(f)$ is the image of the set represented by $f$ under $t$. In
    other words, in modal powerset quantales, the domain and codomain
    operators are simply the images of source and target maps, respectively.

  \item Let $\pi$ be a path in the convolution quantale $Q^{P(C)}$ on
    the path category $P(G)$ on the directed graph $G$ from
    \autoref{ex:convolution}. Then $\Dom(f)(\pi)$ is $\bot$ unless
    $\pi$ is a constant path (of length zero), in which case
    $\Dom(f)(\pi)$ takes all paths $\pi'$ that start in $\pi$, computes
    their values $f(\pi')$ in $Q$, takes the domain elements of these
    values in $Q$ and then computes their sup. If $Q=2$, then
    $\Dom(f)$ computes the sets of sources of paths in the set
    represented by $f$ and $\Cod(f)$ the set of all targets.
  \end{enumerate}
\end{exa}

To define $\Dom$ and $\Cod$ on convolution Kleene algebra, we take the definitions for convolution quantales as the starting point. As the general sups in the definitions of $\Dom$ and $\Cod$ cannot be
expressed with semirings or Kleene algebras, restrictions need to be
imposed. In the tradition of algebra we could consider finitely
supported functions $f:C\to Q$ only or even assume $C$ or $Q$ to be
finite. But this would defeat the purpose of Möbius catoids or Möbius
categories. Instead we consider two different restrictions:
\begin{enumerate}
\item to $S[C]$ or $K[C]$.
\item to Möbius catoids $C$ of \emph{finite valency}~\cite{CalkMPS25},
  that is, the sets $\{x\in C\mid s(x)=e\}$ and
  $\{x \in C\mid t(x)=e\}$ are finite for each $e\in C_0$.
\end{enumerate}

We first explore the first alternative: we define domain and codomain in the modal convolution algebra as for quantales, and the sups are finite because we restrict the convolution algebra to maps in $S[C]$ or $K[C]$. This alternative leads to  trivial algebras of propositions:

\begin{lem}
  \label{lem:dom-one}
  Let $C$ be a catoid and $S$ a modal semiring. Then, for all
  $f\in S[C]$,
\begin{equation*}
  \Dom(f) = \Cod(f) =
  \begin{cases}
    \id_0 & \text{if } f\neq 0,\\
    0   & \text{otherwise}
  \end{cases}
\end{equation*}
and therefore $S[C]_0=\{0,\id_0\}$.
\end{lem}
\begin{proof}
  In any modal semiring, $\dom(1)=1$ and $\cod(1)=1$. Hence, if
  $f\neq 0$, then
\begin{equation*}
  \Dom(f)(x) = \dom(f(x))\cdot[x\in C_0] = [x\in C_0]= \cod(f(x))
  \cdot[x\in C_0]=\Cod(f)(x),
\end{equation*}
because $\dom(f(x))$ and $\cod(f(x))$ either are $1$ and dominate the
sup in the definition of $\Dom(f)(x)$ and $\Cod(f)(x)$, when
$x\in C_0$ for both expressions, or $\delta_{s(y)}(x)$ and
$\delta_{t(y)}(x)$ are $0$ when $x\notin C_0$.

Otherwise, if $f=0$, then trivially $\Dom(0)=\Cod(0)=0$.
\end{proof}

\begin{prop}\label{prop:modal-semiring}
  If $C$ is a finitely $2$-decomposable local catoid and $S$ a modal
  semiring, then $S[C]$ forms a modal semiring with $\Dom$ and $\Cod$
  extended as in \autoref{lem:dom-one} and with $S[C]_0=\{0,\id_0\}$.
\end{prop}
\begin{proof}
  \autoref{prop:conv-dioid-special} shows that $S[C]$ forms a dioid,
  \autoref{lem:dom-one} that $\Dom(f),\Cod(f)\in S[C]$ for all
  $f\in S[C]$ and that $S[C]_0=\{0,\id_0\}$. Further, $Q^C$ forms a
  modal quantale if $C$ is a local catoid and $Q$ a modal
  quantale~\cite[Theorem 7.1]{FahrenbergJSZ23}, and it is routine to
  check that all sups in the proof of this theorem remain finite if
  $C$ is finitely $2$-decomposable. Modal quantales and modal
  semirings have the same axioms. So they hold in particular for the
  $\Dom$ and $\Cod$ maps in $S[C]$.
\end{proof}

While \autoref{prop:modal-semiring} and \autoref{lem:dom-one} show that
$\Dom$ and $\Cod$ can be extended from $S$ and $C$ to $S[C]$,
\autoref{lem:dom-one} also indicates that $\Dom$ and $\Cod$ can be
defined directly on any $S[C]$. The following proposition confirms
this.

\begin{prop}\label{prop:modal-semiring-var}
  If $C$ is a finitely $2$-decomposable local catoid and $S$ a
  dioid, then $S[C]$ forms a modal semiring with $\Dom$ and $\Cod$
  defined by the formulas in \autoref{lem:dom-one}.
\end{prop}
\begin{proof}
  \autoref{prop:conv-dioid-special} shows that $S[C]$ forms a
  dioid, and it follows from \autoref{rem:test-trivial} that
  $S[C]_0=\{0,\id_0\}$. It remains to check that $\Dom$ and $\Cod$
  from \autoref{lem:dom-one} satisfy the domain and codomain axioms.
\begin{itemize}
\item For $f=0$, $\Dom(0)\ast 0 = 0$, and otherwise, for $f\neq 0$,
$\Dom(f)\ast f = \id_0\ast f = f$.

\item For $f=0$ or $g=0$, $\Dom(f\ast \Dom(0)) = 0 = \Dom(f\ast g)$
  and for $f\neq 0\neq g$,
  \begin{equation*}
    \Dom(f\ast \Dom(g))= \Dom(f\ast
    \id_0)=\Dom(f)=\id_0=\Dom(f\ast g),
  \end{equation*}
  because $f\ast g\in S[C]$.

\item It is obvious that $\Dom(f)\le \id_0$ and $\Dom(0)=0$. 

\item If $f=g=0$, then $\Dom(f+g) = 0 = \Dom(f)+\Dom(g)$, and if
  $f\neq 0$ or $g\neq 0$, then
  \begin{equation*}
    \Dom(f+g) = \id_0 = \Dom(f)+\Dom(g),
   \end{equation*}
  because $f+g\in S[C]$ and addition in dioids is idempotent.
\end{itemize}
The proofs for the codomain axioms are dual.
\end{proof}

\begin{cor}\label{cor:modal-ka1}\label{cor:cat-modal-convolution1}
  If $C$ is a local Möbius catoid and $K$ a (modal) Kleene algebra,
  then $K[C]$ forms a modal Kleene algebra with $K[C]_0=\{0,\id_0\}$.
  \end{cor}
  \begin{proof}
    Propositions~\ref{prop:conv-ka-special}, \ref{prop:modal-semiring}
    and \ref{prop:modal-semiring-var} show that $K[C]$ forms a modal
    semiring satisfying $K[C]_0 = \{0,\id_0\}$ and closed under the
    operations of modal Kleene algebra. The result is then immediate
    from \autoref{thm:main-theorem}.
  \end{proof} 

 In the second case, we again define (co)domain as for modal convolution quantales, but we obtain finiteness of the sums $\Dom(f)$
 and $\Cod(f)$ by assuming that $C$ has finite valency.
\begin{prop}\label{thm:modal-semiring}
  If $C$ is a finitely $2$-decomposable local catoid of finite valency
  and $S$ a modal semiring, then $S^C$ forms a modal semiring with
  $\Dom$ and $\Cod$ defined as for quantales.
\end{prop}
\begin{proof}
  It is routine to check that all sups in~\cite[proof of Theorem
  7.1]{FahrenbergJSZ23} remain finite.
\end{proof}
Once again we can combine this result with
\autoref{thm:main-theorem}.
\begin{cor}\label{cor:modal-ka}\label{cor:cat-modal-convolution}
  If $C$ is a local Möbius catoid of finite valency and $K$ a modal
  Kleene algebra, then $K^C$ forms a modal Kleene algebra.
 \end{cor}

\begin{exa}
  Let $K$ be a modal Kleene algebra.
  \begin{enumerate}
  \item For the free monoid and the shuffle catoid on $A^\ast$ from
    Examples~\ref{ex:conv-kas}(1) and (2), the modal structure on
    $K[{A^\ast}]$ is trivial: $\Dom$ and $\Cod$ send each map in
    $K[A^\ast]$ either to the constant $0$ function or to $\id_0$,
    which in this case has codomain $\{\varepsilon\}$.  Yet $A^\ast$
    does not have finite valency.
  \item For the incidence modal Kleene algebra $K[{I_P}]$ on the poset
    $P$ from \autoref{ex:conv-kas}(3), and where each element of
    $I_P$ is finitely $2$-decomposable, the maps $\Dom(f)$ and
    $\Cod(f)$ assign to each unit interval in $I_P$ the weight $1$
    using $f$ and are undefined on all non-unit
    intervals. Alternatively, if every element in the poset $P$ has a
    finite valency (as a graph), and if every interval in $I_P$ is
    finitely $2$-decomposable, then $K^{I_P}$ is a modal Kleene
    algebra. Interval categories have many units and therefore require
    the approach developed in this article. These constructions allow
    enriching interval temporal logics and duration calculi with
    additional modal operators. 

  \item The classical modal semiring is arguably the modal semiring of
    binary relations on a pair groupoid $X\times X$ obtained by
    powerset extension. In this case, the domain and codomain of a
    relation $R\subseteq X\times X$ are
    $\dom(R) = \{x\in X\mid (x,y)\in R \text{ for some } y\in X\}$ and
    $\cod(R) = \{y\in X\mid (x,y)\in R \text{ for some } x\in X\}$,
    and the modal diamond operators satisfy the standard Kripke
    semantics:
    \begin{align*}
      |R\rangle P &=\{x\in X\mid (x,y)\in R \text{ and } y\in P
                    \text{ for some } y\in X\},\\
      \langle R| P &=\{y\in X\mid (x,y)\in R \text{ and } x\in P
      \text{ for some } x\in X\}.
    \end{align*}
    A modal Kleene algebra is obtained taking the
    reflexive-transitive closure $R^\ast = \bigcup_{i\ge 0} R^i$ of a
    relation $R$ as the Kleene star. General modal convolution Kleene
    algebras of $K$-valued relations require Conway's star definition,
    as in \autoref{ex:conv-kas}, and  the underlying pair
    groupoid must be finite to express convolution.
    
  \item For the modal convolution Kleene algebra $K[P(G)]$ on a finite
    graph $G$ from \autoref{ex:conv-kas}(5), the maps $\Dom(f)$ and
    $\Cod(f)$ assign to each constant path in $G_0$ a value using $f$
    and are undefined on all non-unit intervals. For $K^{P(G)}$, note
    that finite valency excludes loops. Hence finite directed acyclic
    graphs $G$ are needed to make $K^{P(G)}$ a modal Kleene algebra.
  \end{enumerate}
\end{exa}

\begin{rem}
  Every modal Kleene algebra $K$ forms a Kleene algebra with tests
  with test algebra $K_0$, in which the second sort is merely a
  distributive lattice. Every model of a modal Kleene algebra is
  therefore a model of such a Kleene algebra with tests. A boolean
  test algebra (or algebra of propositions) can be obtained in modal
  convolution quantales in which the value quantales are
  boolean~\cite{FahrenbergJSZ23}. This means that their underlying
  lattices are boolean algebras. In a similar way one can consider
  (modal) value Kleene algebras where the underlying semiring is a
  boolean semiring. In this case, modal box and diamond operators can
  be defined as De Morgan duals of the diamond operators with respect
  to boolean complementation: for all $\alpha\in K$ and $p\in K_0$,
  \begin{equation*}
    |\alpha ]p= -|\alpha\rangle (-p)\qquad \text{and}
    \qquad [\alpha| p = - \langle \alpha | (-p).
  \end{equation*}

  In program verification, $|\alpha]p$ is known as the \emph{weakest
    liberal precondition operator} of program $\alpha$ and proposition
  $p$; validity of a Hoare triple (or partial correctness
  specification) can be encoded as $p\le |\alpha]q$ or equivalently as
  $\langle \alpha| p \le q$, for $\alpha\in K$ and $p,q\in
  K_0$. Intuitively, $|\alpha]p$ formalises the largest set of program
  states or the most general proposition about program states such
  when program $\alpha$ is executed and terminates, then it must do so
  in states satisfying $p$. The Hoare triple $p\le |\alpha ]q$ thus
  expresses the fact that whenever program $\alpha$ is executed from
  states satisfying $p$ it must end up in states satisfying $q$ upon
  termination.  See~\cite{DesharnaisS11,GomesS16} for more details on
  Hoare logics and predicate transformer algebras based on modal
  Kleene algebra.
\end{rem}

While other definitions of $\Dom$ and $\Cod$ are possible, modal
quantales generally provide a more liberal setting for modal
convolution algebras than modal semiring and Kleene algebras. This is
not only due to the constraints on the function space or the
underlying catoid needed, but also because interesting models such as
weighted relations are not captured by the restriction to Möbius
categories.

\begin{rem}
  In light of \autoref{prop:modal-semiring-var}, one may ask
  whether it suffices to map from any local Möbius catoid $C$ into an
  arbitrary dioid or Kleene algebra $K$ to obtain a modal
  structure on $K^C$. The answer is negative: in the dioid $0<1<a$
  with $a\cdot a=0$, $\dom(a)$ must be $1$ because
  $d(x)=0\iff x=0$ in any modal semiring. Then
    $\dom(a\cdot a)=\dom(0)=0 < 1 = \dom(a) = \dom(a\cdot
  \dom(a))$~\cite{DesharnaisS11}. Hence this dioid cannot be extended
  to a modal semiring and the axiom
  $\dom(x\cdot y)=\dom(x\cdot \dom(y))$ is not available in $K$ to
  derive its analogon in $K^C$.
\end{rem}

%%%%%%%%%%%%%%%%%%%%%%%%%%%%%%%%%%%%%%%%%%%%%%%%%%%%%%%%%%

\section{Concurrent convolution Kleene
  algebras}\label{s:concurrent-convolution-ka}

We now construct concurrent convolution Kleene algebras from Möbius
$2$-catoids with many units, and strict $2$-category in particular.
Concurrent Kleene algebras have been proposed as models for concurrent
systems~\cite{HoareMSW11}. Our construction generalises a previous
result for catoids with a single unit and an ad-hoc grading
condition~\cite{CranchDS21}. In fact, we prove this result for
slightly more general interchange Kleene algebras, as explained below.

A \emph{$2$-catoid}~\cite{CalkMPS25,CranchS24} is a structure
$(C,\odot_0,\odot_1,s_0,t_0,s_1,t_1)$ such that,\\
$(C,\odot_i,s_i,t_i)$, for $i\in
\{0,1\}$, is a catoid and for all $i\neq j$, $i,j\in\{0,1\}$,
\begin{gather*}
s_i\circ s_j = s_j\circ s_i,\qquad s_i\circ t_j=t_j\circ s_i,\qquad
t_i\circ s_j=s_j\circ t_i,\qquad t_i\circ t_j = t_j\circ t_i,\\
s_i(x\odot_j y) \subseteq s_i(x)\odot_j s_i(y)),\qquad t_i(x\odot_j
y)\subseteq  t_i(x)\odot_j t_i(y)),
\end{gather*}
and
\begin{gather*}
   (w \cdot_1 x)\cdot_0 (y\cdot_1 z) \subseteq (w\cdot_0 y) \cdot_1
   (x\cdot_0 z),\\
   s_1\circ s_0 = s_0,\qquad s_1\circ t_0 = t_0,\qquad t_1\circ  s_0=
   s_0,\qquad t_1\circ t_0 = t_0,\\
   s_1(s_1(x)\odot_0 s_1(y)) = s_1(x)\odot_0 s_1(y),\qquad t_1(t_1(x)\odot_0 t_1(y)) = t_1(x)\odot_0 t_1(y).
 \end{gather*}

 A \emph{strict $2$-category}~\cite{MacLane98} is a $2$-catoid that is
 local and functional with respect to the $0$- and the $1$-structure:
 $(C,\odot_i,s_i,t_i)$ is local and functional for $i\in \{0,1\}$.

A \emph{Möbius $2$-catoid} is a $2$-catoid in which the $0$-structure
and $1$-structure are Möbius; a \emph{strict Möbius $2$-category} is a
Möbius $2$-catoid that is a strict $2$-category.

This definition of $2$-catoid from~\cite{CranchS24} extends a
predecessor~\cite{CalkMPS25} by the two final closure axioms.

\begin{rem}
  A structural explanation of $2$-catoids is as follows. We have
  already seen that catoids are relational monoids and hence monoid
  objects in $\mathbf{Rel}$. This generalises to $2$-fold monoid
  objects in $2$-fold monoidal categories~\cite{AguiarM10}, but these
  monoid objects are too strict for modelling $2$-catoids. Instead,
  lax $2$-fold monoid objects can be defined in monoidal
  bicategories. It has been shown in~\cite[Theorem 5.9.5]{CranchS24}
  that $2$-catoids are precisely $2$-fold relational monoids as lax
  $2$-fold monoid objects in the bicategory $\mathbf{Rel}$. In
  particular, the above closure axioms in $2$-catoids arise as
  coherence conditions in $2$-fold relational monoids, while they were
  absent in~\cite{CalkMPS25}. The argument extends to $n$-catoids and
  lax $n$-fold monoid objects in $\mathbf{Rel}$, which appear in the
  next section.
\end{rem}

In every $2$-catoid $C$, the set $C_0$ of fixpoints of $s_0$ and $t_0$
and the set $C_1$ of fixpoints of $s_1$ and $t_1$ satisfy
$C_0\subseteq C_1\subseteq C$. Thus all identities of $0$-composition
remain identities of $1$-composition.

 \begin{rem}\label{remark:globular-shape}
 In a strict $2$-category $C$, the elements of $C_0$ correspond to
 objects or $0$-cells of the category, the elements of $C_1$ to arrows
 or $1$-cells and the elements of $C_2$ to higher cells, more
 specifically $2$-cells. The axioms of strict $2$-categories assemble
 these structures into globular cells
 \begin{equation*}
 \begin{tikzcd}
  s_0(x) \arrow[r, bend left, "s_1(x)", ""{name=U,inner sep=1pt,below}]
  \arrow[r, bend right, "t_1(x)"{below}, ""{name=D,inner sep=1pt}]
  & t_0(x)
  \arrow[Rightarrow, from=U, to=D, "x"]
\end{tikzcd}
\end{equation*}
Strict $2$-categories are therefore known as \emph{globular}
$2$-categories. The $0$ composition and $1$-composition of two composable
$2$-cells $x$ and $y$ can be visualised as
 \begin{equation*}
    \begin{tikzcd}
      s_0(x) \arrow[r, bend left, "s_1(x)", ""{name=U,inner sep=1pt,below}]
      \arrow[r, bend right, "t_1(x)"{below}, ""{name=D,inner sep=1pt}]
      & t_0(x) \arrow[Rightarrow, from=U, to=D, "x"] \arrow[r, bend left, "s_1(y)", ""{name=UU,inner sep=1pt,below}]
      \arrow[r, bend right, "t_1(y)"{below}, ""{name=DD,inner sep=1pt}]
      & t_0(y) \arrow[Rightarrow, from=UU, to=DD, "y"]
    \end{tikzcd}
    \qquad\text{ and }\qquad
    \begin{tikzcd}
      s_0(x) \arrow[r, bend left=70, "s_1(x)", ""{name=U,inner
        sep=3pt,below}]
      \arrow[r, ""{name=M, inner sep=3pt}]\arrow[r, ""{name=M1, inner
        sep=3pt, below}]
      \arrow[r, bend right=70, "t_1(y)"{swap}, ""{name=D,inner sep=3pt}]
      & t_0(y) \arrow[Rightarrow, from=U, to=M, "x"] \arrow[Rightarrow,
      from=M1, to=D, "y"]
    \end{tikzcd}
  \end{equation*}
respectively. Again, the $0$- and $1$-cells in these compositions are
determined by the axioms of strict $2$-categories. A paradigmatic
example of a strict $2$-category is the category $\mathbf{Cat}$ of all
small categories with small categories as $0$-cells, functors as
$1$-cells and natural transformations as $2$-cells. $2$-Catoids still
have the globular cell shape, but the relations defining $0$- and $1$-
compositions are more difficult to visualise. See~\cite{CalkMPS25} for
a discussion.
\end{rem}

There is redundancy in the axioms above.
\begin{lem}[\cite{CranchS24}]
  In any double catoid formed by two catoids $(C,\odot_i,s_i,t_i)$ for
  $i\in \{0,1\}$, the $2$-catoid axioms are derivable from the
  irredundant axioms
  \begin{gather*}
    s_1(x\odot_0 y) \subseteq s_1(x)\odot_0 s_1(y)),\qquad
    t_1(x\odot_0
y)\subseteq  t_1(x)\odot_0 t_1(y)),\\
   (w \cdot_1 x)\cdot_0 (y\cdot_1 z) \subseteq (w\cdot_0 y) \cdot_1
   (x\cdot_0 z),\\
   s_1(s_1(x)\odot_0 s_1(y)) = s_1(x)\odot_0 s_1(y).
\end{gather*}
\end{lem}
Irredundancy of the reduced axiomatisation has been established using
the Isabelle/HOL proof assistant and its SAT solvers~\cite{CalkS24}.

The following definition generalises a previous notion of interchange
semiring~\cite{CranchDS21} to many identities in light of more general
results in~\cite{CranchDS21} and of notions of higher
quantales~\cite{CalkMPS25}.

An \emph{interchange semiring} is a structure
$(S,\cdot_0,\cdot_1,+,0,1_0,1_1)$ such that the $(S,+,\cdot_i,0,1_i)$
are dioids and, for all $\alpha,\beta,\gamma,\delta\in S$,
\begin{equation*}
  (\alpha \cdot_1 \beta)\cdot_0 (\gamma\cdot_1 \delta) \le (\alpha\cdot_0 \gamma) \cdot_1 (\beta\cdot_0
  \delta),\qquad 1_0 \le 1_1.
\end{equation*}
An \emph{interchange Kleene algebra}
$(K,\cdot_0,\cdot_1,+,0,1_0,1_1,(-)^{\ast_0},(-)^{\ast_1})$ is an
interchange semiring formed by two Kleene algebras
$(K,+,\cdot_i,0,1_i,(-)^{\ast_i})$ with $i\in \{0,1\}$.

\begin{thm}[\cite{CranchDS21}]\label{thm:concurrent-semiring}
  If $C$ is a finitely $2$-decomposable $2$-catoid with respect to the
  $0$-structure and $1$-structure and $S$ an interchange semiring,
  then $S^C$ is an interchange semiring.
\end{thm}

In fact, relational $2$-monoids have been used in~\cite{CranchDS21}
instead of $2$-catoids, but the two structures are isomorphic.  The
two closure axioms for $2$-catoids are not needed in this proof.

\begin{cor}\label{cor:ic-ka}\label{cor:ic-convolution}
  If $C$ is a Möbius $2$-catoid and $K$ an interchange Kleene
  algebra, then $K^C$ forms an interchange Kleene algebra.
\end{cor}
\begin{proof}
  This is immediate from \autoref{thm:main-theorem} and
  \autoref{thm:concurrent-semiring}.
\end{proof}

This corollary generalises a theorem from~\cite{CranchDS21} to catoids
and Kleene algebras with several units. It shows how weights can be
added to two-dimensional globular cell structures in a coherent way.

\begin{rem}
  Interchange semirings and Kleene algebras are variants of the
  concurrent semirings and Kleene algebras from concurrency
  theory~\cite{HoareMSW11}: a concurrent semiring is an interchange
  semiring in which $1_0=1_1$ and $\odot_1$ is commutative, and
  likewise for concurrent Kleene algebras. \autoref{cor:ic-convolution}
  adapts immediately: if $C$ is a Möbius $2$-catoid in which $\odot_1$
  is commutative and $K$ a concurrent Kleene algebra, then $K^C$ forms
  a concurrent Kleene algebra. The proof can be obtained by adapting a
  similar result in~\cite{CranchDS21} that describes a commutative
  extension from a catoid and a value quantale to a convolution
  quantale.
\end{rem}

\begin{exa}
  The abstract setting of $2$-catoids and interchange semirings leads
  to instances that are of interest in concurrency theory.
  \begin{enumerate}
 \item  The free monoid structure and the shuffle catoid structure on
  $A^\ast$ interact as a Möbius $2$-catoid with the free monoid
  structure as the $0$-catoid and the shuffle catoid as the
  $1$-catoid. As in Examples~\ref{ex:catoids} (1) and (2), the source
  and target structure is trivial: $C_0=C_1=\{\varepsilon\}$, which
  gives in particular the closure axioms. It is straightforward, but
  somewhat tedious, to check the interchange law
  \begin{equation*}
    (w\| x)\cdot (y\| z)\subseteq (w\cdot y)\| (x\cdot z)
  \end{equation*}
  in $K^{A^\ast}$ using nested structural induction. By
  \autoref{cor:ic-ka}, $K^{A^\ast}$ thus carries an interchange Kleene
  algebra structure whenever $K$ is an interchange Kleene algebra,
  confirming a more direct construction in~\cite{CranchDS21}. This
  example requires the generality of $2$-catoids owing to the
  underlying shuffle catoid.

\item Any class of directed graphs that contains the empty graph and
  is closed under graph join and disjoint union forms a $2$-category
  with graph join as $0$-composition, disjoint union as (commutative)
  $1$-composition and the empty graph as shared unit~\cite[Proposition
  50]{CranchDS21}. If all directed graphs in such a class are finite,
  the $2$-catoid is Möbius and the class of antitone maps in $K^C$
  forms an interchange Kleene algebra whenever $K$ is an interchange
  Kleene algebra, observing that the conditions used
  in~\cite[Corollary 54]{CranchDS21} amount to the fact that the class
  of graphs is Möbius. For classes of finite posets, the operations of
  graph join and disjoint union are known as series and parallel
  composition. The extension results to Kleene algebras carry over.
  In the partial order semantics of concurrency one considers
  isomorphism classes of finite posets whose elements are labelled
  using a finite set. These structures are known as
  \emph{pomsets}. Series and parallel composition extend from finite
  labelled posets to pomsets, yielding again a $2$-category with the
  equivalence class of the empty pomset as a shared
  unit~\cite[Proposition 61]{CranchDS21}. The extension to convolution
  Kleene algebras is as for directed graphs.
 
 \item In higher-dimensional rewriting, a $1$-\emph{computad} (or
  $1$-\emph{polygraph}) is  a graph in the sense of
  \autoref{ex:catoids} and \autoref{rem:quiver}. In rewriting
  theory, $1$-polygraphs are known as \emph{abstract rewriting
    systems}. To construct a $2$-polygraph (or second-order abstract
  rewriting system), one first forms the path category of the
  $1$-polygraph. The vertices of the graph can be seen as a set of
  $0$-generators and the edges as the set of $1$-generators of this
  free category. Then one adds a set of $2$-generators, which form a
  cellular extension of the $1$ polygraph when equipped with elements
  of the $1$-path category as sources and targets. Using a
  $2$-polygraph one can then construct the free $2$-category generated
  by the $0$-, $1$- and $2$-generators as a higher path category. By
  definition, this free category is a $2$-category, hence for any
  $2$-polygraph one can construct a convolution interchange Kleene
  algebra on the path $2$-category it generates. See~\cite{AraBGMMM23}
  for details on higher-dimensional rewriting and polygraphs.
  \end{enumerate}
\end{exa}

In interchange semirings and Kleene algebras, the source and target
structure of the $2$-catoid $C$ imposes the globular structure of $C$.
Yet it is once again collapsed into $\id_0$ and $\id_1$ in $K^C$ and
it only appears in the condition $\id_0\le \id_1$. As previously with
modal semirings and Kleene algebras, we can use domain and codomain
operations in the interchange algebras to make the globular structure
explicit. This is the purpose of the next section, while at the same
time generalising from $2$ to $n$ dimensions.

%%%%%%%%%%%%%%%%%%%%%%%%%%%%%%%%%%%%%%%%%%%%%%%%%%%%%%%%%%%%%

\section{Convolution \texorpdfstring{$n$}{n}-Kleene algebras}\label{s:convolution-n-ka}

Now we consider the construction of $n$-Kleene algebras, which have
been proposed as algebras supporting coherence proofs in
higher-dimensional rewriting~\cite{CalkGMS22}. This allows us to
answer a question in~\cite{CalkMPS25} related to conditions under
which such $n$-Kleene algebras can be obtained from more general
constructions for $n$-quantales. Note that the axiomatisation
introduced in this section differs slightly from that
of~\cite{CalkMPS25}, due to the additional closure axioms on
$n$-catoids, from which $n$-Kleene algebras with additional closure
axioms are obtained.

First, the notion of $2$-catoid in the previous section generalises
readily to a notion of $n$-catoid or $\omega$-catoid, as $n$-catoids
are simply a stack of pairs of $2$-categories.

An \emph{$n$-catoid}~\cite{CalkMPS25,CranchS24} is a structure
$(C,\odot_i,s_i,t_i)_{0\le i <n}$ such that each \\ $(C,\odot_i,s_i,t_i)$
is a catoid, and for all  $i\neq j$, $0\le i,j<n$,
\begin{gather*}
s_i\circ s_j = s_j\circ s_i,\qquad s_i\circ t_j=t_j\circ s_i,\qquad
t_i\circ s_j=s_j\circ t_i,\qquad t_i\circ t_j = t_j\circ t_i,\\
s_i(x\odot_j y) \subseteq s_i(x)\odot_j s_i(y)),\qquad t_i(x\odot_j
y)\subseteq  t_i(x)\odot_j t_i(y)),
\end{gather*}
for all $0\le i<j<n$,
\begin{gather*}
   (w \cdot_j x)\cdot_i (y\cdot_j z) \subseteq (w\cdot_i y) \cdot_j
   (x\cdot_i z),\\
   s_j\circ s_i = s_i,\qquad s_j\circ t_i = t_i,\qquad t_j\circ  s_i=
   s_i,\qquad t_j\circ t_i = t_i,\\
   s_j(s_j(x)\odot_i s_j(y)) = s_j(x)\odot_i s_j(y),\qquad t_j(t_j(x)\odot_i t_j(y)) = t_j(x)\odot_i t_j(y).
 \end{gather*}
 Once again, relative to~\cite{CalkMPS25}, the closure axioms in the
 last line have been added in~\cite{CranchS24}.  This axiomatisation
 can be reduced as for $2$-catoids~\cite{CranchS24}.

 As for $2$-categories, categories of $n$-catoids, with homomorphisms
 preserving each catoid structure, are equivalent to categories of lax
 $n$-fold relational monoids in monoidal bicategories with suitable
 morphisms~\cite{CranchS24}, which justifies the axioms
 in~\cite{CranchS24} from a structural point of view.

 An $n$-catoid is \emph{local} (\emph{functional}) if it is local
 (functional) in each dimension. A \emph{strict $n$-category} is a
 local functional $n$-catoid~\cite{CranchS24}.

A \emph{Möbius $n$-catoid} is an $n$-catoid which is a Möbius catoid
in each dimension $0\le i <n$. A
\emph{strict Möbius $n$-category} is a Möbius $n$-catoid that is a
strict $n$-category.

In $n$-catoids, we obtain a filtration of fixpoint sets $(Q_i)_{0\le
  i<n}$ for the $s_i$ and $t_i$. Moreover, for all $i,j\le k$,
 \begin{equation*}
   s_i(x)\odot_k s_j(y) =
   \begin{cases}
     \{s_i(x)\} & \text{ if } s_i(x)=s_j(y),\\
     \emptyset & \text{ otherwise}.
   \end{cases}
 \end{equation*}
Higher-dimensional compositions of lower-dimensional elements are
therefore trivial.

\begin{rem}
  As in strict $2$-categories, the higher cell structure imposed by
  the axioms of strict $n$-categories is globular. A $3$-cell, for
  instance, can be imagined as a $3$-dimensional globe and its
  upper and lower faces as the upper and lower spherical $2$-cells
  glued together along the two $1$-cells spanning the
  equator. The cells in $n$-catoids satisfy the same relations,
  but the relations on the $n$ compositions and face maps are once
  again weaker than for $n$-categories and more difficult to
  visualise~\cite{CalkMPS25}.
\end{rem}

Next we introduce the corresponding definitions of semiring and Kleene
algebra.  Once again, $n$-semirings and $n$-Kleene algebras are stacks
of pairs of the corresponding $2$-structures.

An \emph{$n$-semiring} is a structure
$(S,+,0,\cdot_i,1_i,\dom_i,\cod_i)_{0\le i < n}$ such that the
$(S,+,0,\cdot_i,1_i,\dom_i,\cod_i)$ are modal semirings and the
structures interact as follows:
\begin{itemize}
 \item  for all $i\neq j$,
  \begin{equation*}
    \dom_i(\alpha\cdot_j \beta) \le  \dom_i(\alpha)\cdot_j
    \dom_i(\beta)\qquad\text{ and }\qquad
 \cod_i(\alpha\cdot_j \beta) \le \cod_i(\alpha)\cdot_j
 \cod_i(\beta),
\end{equation*}
\item for all $i<j$
\begin{gather*}
  (\alpha\cdot_j \beta) \cdot_i (\gamma\cdot_j \delta) \le (\alpha \cdot_i \gamma)\cdot_j (\beta\cdot_i
  \delta),\\
  \dom_j(\dom_i(\alpha)) = \dom_i(\alpha),\\
  \dom_j(\dom_j(\alpha)\cdot_i\dom_j(\beta)) =
  \dom_j(\alpha)\cdot_i\dom_j(\beta),\qquad \cod_j(\cod_j(\alpha)\cdot_i\cod_j(\beta)) = \cod_j(\alpha)\cdot_i\cod_j(\beta).
\end{gather*}
\end{itemize}

Relative to a previous axiomatisation~\cite{CalkMPS25} we have added
the two closure axioms in the last line. These are irredundant, see
\autoref{A:independence} for details.

 As for $n$-catoids, we obtain a filtration $(S_i)_{0\le i<n}$ of the
 sets of fixpoints of $\dom_i$ and $\cod_i$. Each $S_i$ forms a
 bounded distributive lattice embedded into the bounded distributive
 lattice $S_{i+1}$ if it exists.

 An \emph{$n$-Kleene algebra}~\cite{CalkMPS25} is an $n$-semiring $K$
equipped with Kleene stars $(-)^{\ast_i}:K\to K$ that satisfy the
usual star unfold and star induction axioms for all $0\le i< j <n$ and
\begin{gather*}
    \dom_i (x) \cdot_i y^{\ast_j} \le (\dom_i (x)\cdot_i
  y)^{\ast_j},\qquad x^{\ast_j}\cdot_i\cod_i (y) \le
  (x\cdot_i \cod_i(y))^{\ast_j}.
\end{gather*}

The two additional star axioms are motivated by applications in
higher-dimensional rewriting~\cite{CalkGMS22}.

Constructing convolution $n$-semirings and $n$-Kleene algebras
requires again restrictions on domain and codomain. As in the
$1$-dimensional case, we could define function spaces $S[C]$ and
$K[C]$, for $S$ an $n$-dioid and $K$ an $n$-Kleene algebra. But this would
trivialise the weights of cells in all dimension below $n$. In a
strict $2$-category of paths, for instance, $C_0$ would model
vertices, $C_1$ paths between vertices and $C_2$ higher cells between
paths, and the approach outlined would assign the same weight to all
paths between two vertices. We therefore restrict our attention to the
alternative finite-valency-based approach.

An $n$-catoid $C$ has \emph{finite valency} if each of the underlying
catoids $C_i$ has this property.

\begin{thm}\label{thm:n-semiring}
  If $C$ is a finitely $2$-decomposable local $n$-catoid of finite valency
  and $S$ an $n$-semiring, then $S^C$ forms an $n$-semiring.
\end{thm}
\begin{proof}
  It is shown in~\cite{CalkMPS25}  that $Q^C$ satisfies all $n$-quantale axioms except the
  closure axioms. For finitely $2$-decomposable local
  $n$-catoids of finite valency, all sups in this proof remain
  finite. It remains to check the closure axioms.
\begin{align*}
&\Dom_j(\Dom_j(f)\ast_i\Dom_j(g))(x)\\
&= \sum_y\dom_j\left(\sum_{u,v}\left(\sum_a
                                      \dom_j(f(a))\delta_{s_j(a)}(u)\right)\cdot_i\left(\sum_b\dom_j(g(b))\delta_{s_j(b)}(v)\right)[y\in
                                         u\odot_iv]\right)\delta_{s_j(y)}(x)\\
&= \sum_{y,u,v,a,b}\dom_j(
                                      \dom_j(f(a))
                                                                                 \cdot_i\dom_j(g(b)))\delta_{s_j(a)}(u)\delta_{s_j(b)}(v)[y\in u\odot_iv]\delta_{s_j(y)}(x)\\
&= \sum_{u,v,a,b}\dom_j(
                                      \dom_j(f(a))
                                                                                                                                                                                \cdot_i\dom_j(g(b)))\delta_{s_j(a)}(u)\delta_{s_j(b)}(v)[x\in s_j(u\odot_iv)]\\
&= \sum_{u,v,a,b}\dom_j(f(a))
                                                                                                                                                                                                                                                                  \cdot_i\dom_j(g(b))\delta_{s_j(a)}(u)\delta_{s_j(b)}(v)[x\in
                                                                                                                                                                                                                                                                  u\odot_iv]\\
&= \sum_{u,v}\left(\sum_a\dom_j(f(a)) \delta_{s_j(a)}(u)\right)
                                                                                                                                                                                                                                                                                 \cdot_i\left(\sum_b\dom_j(g(b))\delta_{s_j(b)}(v)\right)[x\in u\odot_iv]\\
&=(\Dom_j(f)\ast_i\Dom_j(g))(x).
\end{align*}
The first step unfolds the definition of $\Dom_j$ and $\ast_i$, the
second applies distributivity laws, in particular for $\dom_j$. The
third step shifts the composition constraint from $y$ to $x$, the
fourth step uses the closure axiom on $s_j$. The fifth step uses
distributivity laws and the final step again the definitions of
$\Dom_j$ and $\ast_i$.

The proof of the closure axiom for $\Cod_j$ and $\ast_i$ is dual.
\end{proof}

Extending this theorem to local Möbius $n$-catoids and
$n$-Kleene algebras requires two technical lemmas.

\begin{lem}\label{lem:Dom-1}
  Let $C$ be a finitely 2-decomposable local $n$-catoid of finite
  valency and $S$ an $n$-semiring. Then, for all $f:C\to S$ and $x\in
  C$,
  \begin{equation*}
    \Dom_i(f)(x) \le \Dom_i(f)(x) \cdot_j \Dom_i(f)(x).
  \end{equation*}
\end{lem}
\begin{proof}
  \begin{align*}
    &\Dom_i(f)(x) \cdot_j \Dom_i(f)(x)\\
    &=\sum_{y,y'\in C} \dom_i(f(y))\cdot_j\dom_i(f(y'))
      \delta_{s_i(y)}(x)\delta_{s_i(y')}(x)\\
    &=\sum_{y} \dom_i(f(y))\cdot_j\dom_i(f(y))
      \delta_{s_i(y)}(x)+\sum_{y\neq y'} \dom_i(f(y))\cdot_j\dom_i(f(y'))
      \delta_{s_i(y)}(x)\delta_{s_i(y')}(x)\\
    &\ge \sum_{y} \dom_i(f(y)) \delta_{s_i(y)}(x)\\
    &= \Dom_i(f)(x).
  \end{align*}
In the penultimate step, $\dom_i(x)\cdot_j\dom_i(x)=\dom_i(x)$ holds
in every $n$-quantale~\cite[Lemma 7.7(1)]{CalkMPS25}, and hence in
every $n$-semiring.
\end{proof}

\begin{lem}\label{lem:Dom-2}
  Let $C$ be a finitely 2-decomposable local catoid of finite valency
  and $S$ a semiring. Then, for all $f:C\to S$ and $x\in C$,
  \begin{equation*}
    (\Dom(f)\ast g)(x)= \Dom (f)(s(x))\cdot g(x).
    \end{equation*}
  \end{lem}
  \begin{proof}
    \begin{align*}
       (\Dom(f)\ast g)(x)
    &= \sum_{y,z\in C} \Dom(f)(y)\cdot g(z)[x\in y\odot
      z]\\
    &= \sum_{z\in C} \Dom(f)(s(x))\cdot g(z)[x\in s(x)\odot
      z]\\
    &= \sum_{z\in C} \Dom(f)(s(z))\cdot g(z)[x\in s(z)\odot
      z]\\
    &= \sum_{z\in C} \Dom(f)(s(z))\cdot g(z)[x = z]\\
      &= \Dom (f)(s(x))\cdot g(x),
    \end{align*}
      because only $\Dom(f)(y)$ with $y=s(x)$ contributes to the
  sum.
  \end{proof}

\begin{thm}\label{thm:n-ka}
  If $C$ is a local Möbius $n$-catoid of finite valency and $K$ an
  $n$-Kleene algebra, then $K^C$ forms an $n$-Kleene algebra.
\end{thm}

\begin{proof}
  Relative to Theorems~\ref{thm:main-theorem} and \ref{thm:n-semiring}
  it remains to derive the two star axioms mentioning domain and
  codomain in $K^C$. We only show
  $ (\Dom_i(f)\ast_ig^{\ast_j})\le (\Dom_i(f)\ast_ig)^{\ast_j}$ for $0\leq i < j < n$, the
  proof for $\Cod_i$ being dual.  We proceed by induction on $\ell(x)$
  in
  \begin{align*}
    (\Dom_i(f)\ast_ig^{\ast_j})(x) = \Dom_i(f)(s_i(x))\cdot_i g^{\ast_j}(x),
  \end{align*}
which is obtained using \autoref{lem:Dom-2}.
   In the base case, if $x= s_i(x)$, then
 \begin{align*}
   (\Dom_i(f)\ast_ig^{\ast_j})(x)
   &=\Dom_i(f)(x)\cdot_i g^{\ast_j}(x)\\
   &=\Dom_i(f)(x)\cdot_i g^{\ast_j}(s_j(x))\\
   &= \Dom_i(f)(x)\cdot_i g(x)^{\ast_j}\\
   &\le (\Dom_i(f)(x)\cdot_i g(x))^{\ast_j}\\
   &= ((\Dom_i(f)\ast_i g)(x))^{\ast_j}\\
   &= (\Dom_i(f)\ast_i g)^{\ast_j}(x),
 \end{align*}
 using the domain axiom in $K$ in the fourth step. Also, in the second
 step. $x=s_j(x)$ because $x=s_i(x)$ by assumption and
 $s_i(x)=s_j(s_i(x))$ is immediate from the $n$-catoid axioms.

 In the induction step, we abbreviate
  $\Sigma_\ast(x) =   \sum_{y,z\in C} g(y)\cdot_j g^{\ast_j}(z)[x\in y\odot_j
        z,y\notin s_j(x)]$ and $D_\ast(x)= (\Dom_i(f)\ast_i g) ^{\ast_j}(s_j(x))$.
 \begin{align*}
  & (\Dom_i(f)\ast_ig^{\ast_j})(x)\\
   &= \Dom_i(f)(s_i(x))\cdot_i g^{\ast_j}(x)\\
   & =\Dom_i(f)(s_i(x))\cdot_i
     \left(g(s_j(x))^{\ast_j}\cdot_j \Sigma_\ast(x)\right)\\
      & \le (\Dom_i(f)(s_i(x))\cdot_j\Dom_i(f)(s_i(x)))\cdot_i
        \left(g(s_j(x))^{\ast_j}\cdot_j \Sigma_\ast(x)\right)\\
   & \le (\Dom_i(f)(s_i(x))\cdot_i g(s_j(x)) ) ^{\ast_j}\cdot_j
     \left(\Dom_i(f)(s_i(x))\cdot_i \Sigma_\ast(x)\right)\\
      & = (\Dom_i(f)(s_i(s_j(x)))\cdot_i g(s_j(x)) ) ^{\ast_j}\cdot_j
     \left(\Dom_i(f)(s_i(x))\cdot_i
       \Sigma_\ast(x)\right)\\
   & = (\Dom_i(f)\ast_i g) ^{\ast_j}(s_j(x))\cdot_j
            \sum_{y,z\in C} \Dom_i(f)(s_i(x))\cdot_i  (g(y)\cdot_j g^{\ast_j}(z))[x\in y\odot_j
     z,y\notin s_j(x)]\\
   & = D_\ast(x)\cdot_j
            \sum_{y,z\in C} (\Dom_i(f)(s_i(x))\cdot_j \Dom_i(f)(s_i(x)))\cdot_i  (g(y)\cdot_j g^{\ast_j}(z))[x\in y\odot_j
     z,y\notin s_j(x)]\\
  & \le D_\ast(x)\cdot_j
            \sum_{y,z\in C} (\Dom_i(f)(s_i(y))\cdot_i g(y))\cdot_j  (\Dom_i(f)(s_i(z))\cdot_i g^{\ast_j}(z))[x\in y\odot_j
    z,y\notin s_j(x)]\\
 & = (\Dom_i(f)\ast_i g) ^{\ast_j}(s_j(x))\cdot_j
            \sum_{y,z\in C} (\Dom_i(f)\ast_i g)(y)\cdot_j  ((\Dom_i(f)\ast_i g ^{\ast_j})(z))[x\in y\odot_j
   z,y\notin s_j(x)]\\
 & \le (\Dom_i(f)\ast_i g) ^{\ast_j}(s_j(x))\cdot_j
            \sum_{y,z\in C} (\Dom_i(f)\ast_i g)(y)\cdot_j  ((\Dom_i(f)\ast_i g) ^{\ast_j} (z))[x\in y\odot_j
   z,y\notin s_j(x)]\\
   &=  (\Dom_i(f)\ast_ig)^{\ast_j}(x).
 \end{align*}
The first three steps prepare for the first application of
 interchange, using \autoref{lem:Dom-1}. The
 constraints on interchange are trivial. In the fourth step we apply the interchange law, and also use the base case. In the fifth and sixth step we use
 \autoref{lem:Dom-2} to rewrite the first factor and the definition
 of the star to unfold the second one. The following three steps lead
 to the second application of interchange, using again
 \autoref{lem:Dom-1}. After applying interchange in the eighth step,
 we use $s_i(x)=s_i(y)=s_i(z )$, which follows from the constraint
 $x\in y\odot_j z$ and
 $x \in s_i(x)\odot_i(y\odot_j z) = (s_i(x)\odot_j s_i(x))\odot_i
 (y\odot_j z)\subseteq (s_i(x)\odot_i y) \odot_j (s_i(x)\odot z)$.  In
 the ninth step, we rewrite again some factors using
 \autoref{lem:Dom-2}. In the tenth, we apply the induction
 hypothesis to the term depending on $z$, which has smaller length
 than $x$ because $y\notin s_j(x)$. In the final step we apply again
 the definition of the star. 
\end{proof}

\begin{cor}\label{cor:cat-ka-convolution}
  If $C$ is a Möbius $n$-category of finite valency and $K$ an
  $n$-Kleene algebra, then $K^C$ forms an $n$-Kleene algebra.
  \end{cor}

  Higher convolution Kleene algebras thus allow us to assign weights
  to cells of $n$-catoids and strict $n$-categories in a coherent way.

\begin{exa}
  Strict $n$-categories have found applications in higher-dimensional
  rewriting~\cite{AraBGMMM23}. Recently, slightly different $n$-Kleene
  algebras have been used to prove basic rewriting properties such as
  coherent Newman's lemmas or coherent Church-Rosser theorems in this
  setting~\cite{CalkGMS22}. The results in this section provide a
  systematic construction of $n$-Kleene algebras from strict
  $n$-categories, in particular from free strict $n$-categories
  generated by computads or polygraphs, which correspond to
  higher-dimensional rewrite systems. As such path categories have
  many units, the framework provided by \autoref{thm:main-theorem} is
  needed for constructing convolution $n$-Kleene algebras over
  them. In the context of polygraphs, however, the finite valency
  restriction used for constructing $\Dom_i$ and $\Cod_i$ seem rather
  severe. They exclude polygraphs generating cyclic paths in any
  dimension. In practice, however, higher-dimensional rewriting
  systems are often assumed to be noetherian, which rules out infinite
  rewriting paths.
\end{exa}

\section{\texorpdfstring{$\ast$}{Star}-continuous convolution Kleene algebras}
\label{A:star}

In this section we show that the recursive definition \eqref{eq:star}
of the Kleene star allows defining $\ast$-contiuous convolution Kleene
algebras from Möbius catoids and $\ast$-continuous value Kleene
algebras.  Conceptually, $\ast$-continuous Kleene algebras are
situated between Kleene algebras and quantales, the differences being
that in $\ast$-continuous Kleene algebras only the sups needed for
defining Kleene stars are assumed. In particular, rational languages
are models of Kleene algebras and $\ast$-continuous Kleene algebras,
but not of quantales, while languages are models of all three
algebras.

Formally, a \emph{$\ast$-continuous Kleene algebra}~\cite{Kozen94} is
a dioid $K$
equipped with a star operation such that, for $\beta^0 = 1$ and
$\beta^{i+1} = \beta\cdot \beta^i$, 
\begin{equation*}
  (\forall i\in \mathbb{N}.\ \alpha\cdot \beta^i \cdot \gamma \le
  \delta) \iff \alpha \cdot \beta^\ast \cdot \gamma \le \delta, 
\end{equation*}

Note that
$(\forall i\in \mathbb{N}.\ \alpha\cdot \beta^i \cdot \gamma \le
\delta) \iff \bigvee_{i \in \mathbb{N}} \alpha\cdot \beta^i\cdot\gamma
\le \delta$, so that the above definition means that, for all
$\alpha,\beta,\gamma\in K$, the sup
$\bigvee_{i \ge 0} \alpha\cdot \beta^i\cdot \gamma$ of the
$\alpha\cdot \beta^i\cdot \gamma$ exists and is equal to
$\alpha\cdot \beta^\ast\cdot\gamma$.  Henceforth we simply write
\begin{equation*}
  \bigvee_{i \ge 0} \alpha\cdot \beta^i\cdot \gamma = \alpha\cdot
  \beta^\ast\cdot\gamma
\end{equation*}
in calculations, where $\bigvee$ indicates an arbitrary sup. The star axioms of Kleene algebra are derivable from
the $\ast$-continuous axiom schema; all $\ast$-continuous Kleene
algebras are Kleene algebras. The above axiom also entails the
quantalic definition $\alpha^\ast = \bigvee_{i\ge 0} \alpha^i$ of the
star and the quantalic sup-preservation laws restricted to sups of
powers.

\begin{thm}
  \label{thm:star-cont-ka-star}
  Let $C$ be a Möbius catoid and $K$ a
  $\ast$-continuous Kleene algebra. Then $K^C$ forms a $\ast$-continuous
  Kleene algebra with star given by \eqref{eq:star}.
\end{thm}

\begin{proof}
  We know that $K$ is a Kleene algebra, hence so is $K^C$ by
  \autoref{thm:main-theorem}. We need to show that
  $fg^i h \le k$ holds for all
  $i\in\mathbb{N}$ if and only if $fg^\ast h \le k$.

  Suppose $fg^i h \le k$ holds for all
  $i\in\mathbb{N}$, that is, for all $x\in C$ and $i\in \mathbb{N}$,
  we have that 
$ 
    (fg^i h)(x)
  =\sum_{w,y,z\in C}f(w)g^i(y)h(z)[x\in wyz]\le k(x)
$.
We show,  for all $x\in
C$,  that
$
  (fg^\ast h)(x) =
  \sum_{w,y,z\in C}f(w)g^\ast(y)h(z)[x\in wyz]\le k(x)$ by induction on $\ell(y)$.

  If $y\in C_0$, then
  \begin{align*}
    \sum_{w,y,z\in C}f(w)g^\ast(y) h(z)[x\in wyz]
    &= \sum_{w,y,z\in C}f(w)g(y)^\ast h(z)[x\in wyz]\\
    &=\sum_{w,y,z\in C}\bigvee_{i\ge 0} f(w)g(y)^i h(z)[x\in wyz]\\
    &=\bigvee_{i\ge 0}\sum_ {w,y,z}f(w)g^i(y)h(z)[x\in wyz]\\
    &=\bigvee_{i\ge 0}(fg^ih)(x)\\
    &\le k(x)
  \end{align*}
  by indecomposability of $x$, which implies $g(y)^i=g^i(y)$, \eqref{eq:star},
  $\ast$-continuity in $K$ and the assumption. The last step uses in
  particular idempotency of addition. 

  If $y\in C_1$, then, writing $I(x,y,z)$ for $[x\in yz,y\neq s(x)]$
  as in \autoref{lem:conv-star-prop},
  \begin{align*}
    &\sum_{w,y,z\in C}f(w)g^\ast(y) h(z)[x\in wyz]\\
    % & = \sum_{w,y,z\in C}f(w)g(s(y))^\ast \left(\sum_{a,b\in C}
    %   g(a)g^\ast(b)I(y,a,b)\right)h(z)[x\in wyz]\\
       & = \sum_{w,y,z\in C}\sum_{a,b\in C} f(w)g(s(y))^\ast
         g(a)g^\ast(b) h(z) I(y,a,b)[x\in wyz]\\
    % & = \sum_{w,y,z\in C}\sum_{a,b\in C} \bigvee_{m\ge 0}
    %   f(w)g(s(y))^\ast  g(a)g^m(b)h(z)I(y,a,b)[x\in wyz]\\
       & = \sum_{w,y,z\in C}\sum_{a,b\in C} \bigvee_{m\ge 0}
         \bigvee_{i\ge 0}f(w)g(s(y))^i  g(a)g^m(b)h(z)I(y,a,b)[x\in
         wyz]\\
      & = \sum_{w,y,z\in C}\sum_{a,b\in C} \bigvee_{i\ge 0}
         \bigvee_{m\ge 0}f(w)g(s(y))^i  g(a)g^m(b)h(z)I(y,a,b)[x\in
        wyz]\\
     & = \sum_{w,y,z\in C}\sum_{a,b\in C} \bigvee_{i\ge 0}
          \bigvee_{n> i}f(w)g(s(y))^i
       g(a)g^{n-1-i}(b)h(z)I(y,a,b)[x\in wyz]\\
    & = \sum_{w,y,z\in C}\sum_{a,b\in C} \bigvee_{n\ge 1}
          \sum_{i=0}^{n-1}f(w)g(s(y))^i
          g(a)g^{n-1-i}(b)h(z)I(y,a,b)[x\in wyz]\\
          %  & = \sum_{w,y,z\in C}\bigvee_{n\ge 1}\sum_{a,b\in C} 
          % \sum_{i=0}^{n-1}f(w)g(s(y))^i
          %    g(a)g^{n-1-i}(b)h(z)I(y,a,b)[x\in wyz]\\
     & = \sum_{w,y,z\in C}\bigvee_{n\ge 1}f(w)\left(\sum_{a,b\in C} 
          \sum_{i=0}^{n-1}g(s(y))^i
       g(a)g^{n-1-i}(b)I(y,a,b)\right)h(z)[x\in wyz]\\
    & = \sum_{w,y,z\in C}\bigvee_{n\ge 1}f(w)g^n(y)h(z)[x\in wyz]\\
& = \sum_{w,y,z\in C}\left(f(w)\id_0(y)h(z)+\bigvee_{n\ge 1}f(w)
                                                                      g^n(y)h(z)\right)[x\in wyz]\\
    & = \sum_{w,y,z\in C}\bigvee_{n\ge 0}f(w) g^n(y)h(z)[x\in wyz]\\
       & = \bigvee_{n\ge 0}\sum_{w,y,z\in C} f(w) g^n(y)h(z)[x\in
         wyz]\\
    & = \bigvee_{n\ge 0}(fg^n h)(x)\\
    &\le k(x).
  \end{align*}

  In the first step we use \eqref{eq:star} to expand the star
  recursively and apply distributivity laws in $K$. In the second step
  we apply the induction hypothesis, replacing Kleene stars by sups of
  powers. In the next five step we rearrange sums to apply
  \eqref{eq:power-exact} from \autoref{lem:conv-star-prop} in step
  seven. In particular, we use the constraint $0\le i<n$ to swap the
  sums on $i$ and $n$ and then set $m=n-1-i$. After some further
  rearrangements of sums and folding back convolutions, we use the
  assumption in the last step.

  Conversely, suppose $fg^\ast h \le k$. It follows from Kleene
  algebra in $K^C$ that $g^i\le g^\ast$ for all $i$, hence
  $fg^i h\le fg^\ast h$ and therefore $fg^i h\le k$ for all $i$ by
  transitivity.
\end{proof}

%%%%%%%%%%%%%%%%%%%%%%%%%%%%%%%%%%%%%%%%%%%%%%%%%%%%%%%%%%

\section{Convolution Conway semirings}\label{s:conway-semirings}

In this section we extend the convolution algebra construction for
from Kleene algebras in \autoref{s:convolution-kas} to Conway semirings;
see \autoref{s:conv-algs} for their axioms. These are standard in
language theory, in the context of weighted automata and formal power
series~\cite{DrosteK09}. They have been widely studied in the
literature for several decades. Bloom and Ésik have shown that
convolution algebras from a monoid to a Conway semiring form a Conway
semiring~\cite{BloomE93}. A generalisation to Möbius catoids and
categories seems worthwhile.

\begin{thm}\label{thm:conway-semirings}
  If $C$ is a Möbius catoid and $S$ a Conway semiring, then $S^C$ can
  be equipped with a Conway semiring structure with star defined as in
  \eqref{eq:star}.
\end{thm}

\begin{proof}
  \autoref{thm:semiring} implies that $S^C$ forms a dioid. We need to
  check the star axioms
\begin{equation*}
\id_0 + f \ast f^\ast = f^\ast,\qquad\id_0+ f^\ast \ast f=
  f^\ast,\qquad (f+ g)^\ast = (f^\ast \ast g)^\ast \ast f^\ast,\qquad f \ast (f\ast g)^\ast = (f \ast g)^\ast f.
\end{equation*}

The left star unfold axiom $\id_0 + f \ast f^\ast =
f^\ast$ follows immediately from the proof of of this axiom in
\autoref{thm:main-theorem}, noting that idempotency of addition is
not used.

The proof right star unfold axiom $\id_0+ f^\ast \ast f= f^\ast$ is
similar, using \autoref{lem:starswap} instead of the definition of
the star to unfold $f^\ast$.

For $(f+ g)^\ast= (f^\ast \ast g)^\ast \ast f^\ast$, we
abbreviate $\lambda = (f + g)^\ast$ and $\rho=(f^\ast g)^\ast f^\ast$
and verify $\lambda(x)=\rho(x)$ by induction on $\ell(x)$.  If
$\ell(x)=0$, then
\begin{align*}
  \lambda(x)=(f(x)+g(x))^\ast
  =(f(x)^\ast g(x))^\ast f(x)^\ast
  =(f^\ast g (x))^\ast f^\ast (x)
  =(f^\ast g)^\ast (x) f^\ast (x)=\rho (x).
\end{align*}
Otherwise, if $\ell(x)> 0$, then
\begin{align*}
  \rho(x)
  &=(f^\ast g)^\ast(s(x))f^\ast(x) + \sum_{y,z}(f^\ast
    g)^\ast(y)f^\ast(z)[x\in yz,y\neq s(x)]\\
  &=(f^\ast g)^\ast(s(x))f^\ast(x) + (f^\ast
    g)^\ast(s(x))\sum_{u,v,z}(f^\ast g)(u)(f^\ast
    g)^\ast(v) f^\ast(z)[x\in uvz,u\neq s(x)]\\
  &=(f^\ast g)^\ast(s(x))f^\ast(x) + (f^\ast
    g)^\ast(s(x))\sum_{u,w}(f^\ast g)(u)\rho(w)[x\in uw,u\neq
    s(x)]\\
  &=(f^\ast g)^\ast(s(x))f^\ast(x) + (f^\ast
    g)^\ast(s(x))\sum_{u_1,u_2,z}f^\ast(u_1)g(u_2)\rho(z)[x\in u_1u_2z,u_1u_2\neq s(x)]\\
  &=(f^\ast g)^\ast(s(x))f^\ast(x) + (f^\ast
    g)^\ast(s(x))\sum_{u_1,u_2, z}f^\ast(u_1)
    g(u_2)\rho(z)[x\in u_1u_2z, u_1\neq s(x)]\\
  &\quad + (f^\ast g)^\ast(s(x)) f^\ast(s(x))\sum_{y,z}
    g(y)\rho(z)[x\in yz, y\neq s(x)]\\
  &=\rho(s(x))\sum_{y,z}f(y)f^\ast(z)[x\in yz,y\neq s(x)]\\
  &\quad+ \rho(s(x))\sum_{w_1,w_2,u_2,z}f(w_1)f^\ast(w_2)
    g(u_2)\rho(z)[x\in w_1w_2u_2z, w_1\neq s(x)]
    \\
  &\quad + \rho(s(x)) \sum_{y,z} g(y)\rho(z)[x\in yz, y\neq s(x)]\\
  &= \rho(s(x)) \sum_{yz} f(y)(id_0+(f^*g)(f^*g)^*) f^*(z)[x\in yz,y\neq s(x)] \\
  &\quad+ \rho(s(x))\sum_{y,z}
  g(y)\rho(z)[x\in yz,y\neq s(x)]\\
    &=\rho(s(x)) \sum_{yz} f(y)\rho(z)[x\in yz,y\neq s(x)] + \rho(s(x))\sum_{y,z}
      g(y)\rho(z)[x\in yz,y\neq s(x)]\\
  &=\lambda(s(x))\sum_{y,z} (f+g)(y)\lambda(z)[x\in yz,y\neq s(x)]\\
  &=\lambda(x).
\end{align*}
In the fifth step we decompose the second summand according to
$u_1\neq s(x)$ and $u_1=s(x)$ in the constraint $u_1u_1\neq s(x)$. In the sixth step we replace $f^*(x)$ and $f^*(u_1)$ with the definition of the star. In the penultimate step we use the base case
and the induction hypothesis for $\ell(z)< \ell(x)$ to replace $\rho$
with $\lambda$.

Finally, for $f \ast (f\ast g)^\ast = (f \ast g)^\ast f$, we
abbreviate $\lambda = f(gf)^\ast$ and $\rho = (fg)^\ast f$ and prove
$\lambda(x)=\rho(x)$ again by induction on $\ell(x)$. If $\ell(x)=0$, then
\begin{align*}
  \rho(x)&= (fg)^\ast(x)f(x) \\
  &=(f(x)g(x))^\ast f(x)\\ &=f(x)(g(x)f(x))^\ast \\ &=f(x)(gf(x))^\ast \\& =f(x)(gf)^\ast(x)=\lambda(x).
\end{align*}
Otherwise, if $\ell(x)>0$, then
\begin{align*}
  \lambda(x)
 % &= \sum_{y,z} f(y)(gf)^\ast(z)[x\in yz]\\
   &= f(s(x)) (gf)^\ast(x) + \sum_{y,z} f(y)(gf)^\ast(z)[x\in yz,y\neq
     s(x)]\\
   &= f(s(x))(gf)^\ast(s(x))\sum_{y,z}(gf)(y)(gf)^\ast(z)[x\in yz,y\neq s(x)] \\
   &\quad+ \sum_{y,z} f(y)(gf)^\ast(z)[x\in yz,y\neq
     s(x)]\\
   &=\lambda(s(x))\sum_{u,v,z}g(u)f(v)(gf)^\ast(z)[x\in uvz,uv\neq s(x)]
   \\ &\quad+\sum_{y,z} f(y)(gf)^\ast(z)[x\in yz,y\neq
     s(x)]\\
   &=\rho(s(x)) g(s(x))\sum_{v,z}f(v)(gf)^\ast(z)[x\in vz,v\neq s(x)]
    \\&\quad+\lambda(s(x))\sum_{u,v,z}g(u)f(v)(gf)^\ast(z)[x\in uvz,u\neq s(x)] \\
    &\quad+ \sum_{y,z} f(y)(gf)^\ast(z)[x\in yz,y\neq s(x)]\\
   &=(id_0+(fg)^\ast(fg))(s(x))\sum_{y,z}f(y)(gf)^\ast(z)[x\in yz,y\neq s(x)]\\
     &\quad+\rho(s(x))\sum_{u,z}g(u)\lambda(z)[x\in uz,u\neq s(x)]\\
     &=(fg)^\ast(s(x))\sum_{y,z}f(y)(gf)^\ast(z)[x\in yz,y\neq s(x)]
     \\
     &\quad+\rho(s(x))\sum_{u,z}g(u)\lambda(z)[x\in uz,u\neq s(x)]\\
     &=(fg)^\ast(s(x))\sum_{y,z}f(y)(id_0+(gf)(gf)^\ast)(z)[x\in yz,y\neq s(x)]
     \\ &\quad+\rho(s(x))\sum_{u,z}g(u)\lambda(z)[x\in uz,u\neq s(x)]\\
   &=
    (fg)^\ast(s(x))f(x)+\rho(s(x))\sum_{u,z}g(u)\rho(z)[x\in
     uz,u\neq s(x)]\\
  &\quad +(fg)^\ast(s(x))\sum_{u,v,z}f(u)g(v)\rho(z)[x\in
    uvz,u\neq s(x)]\\
    &=
    (fg)^\ast(s(x))f(x)+(fg)^\ast(s(x))\sum_{u,v,z}f(u)g(v)\rho(z)[x\in
      uvz,uv\neq s(x)]\\
  &=\rho(x).
\end{align*}
In the third step we use the base case to replace $\lambda$ with
$\rho$. In the fourth we decompose the first summand according to
$u\neq s(x)$ and $u=s(x)$ in the constraint $uv\neq s(x)$.
In the fifth  we combine the first and third summand. In the eighth we
use the induction hypothesis with $\ell(z)<\ell(x)$ to replace
$\lambda$ with $\rho$, and distributivity. In the penultimate step we
combine the two cases of $u\neq s(x)$ and $u= s(x)$ in the
decomposition $x\in uvz$ to obtain the constraint $uv\neq s(x)$ and
one single sum.
\end{proof}

The results in
Sections~\ref{s:modal-convolution-ka}-\ref{s:convolution-n-ka} can be
adapted to Conway semirings, but we leave this as future work.  Many
examples of semirings with non-trivial stars from the literature have
an idempotent addition, hence overall our results for Kleene algebras
seem sufficiently interesting in practice.

%%%%%%%%%%%%%%%%%%%%%%%%%%%%%%%%%%%%%%%%%%%%%%%%%%%%%%%

\section{The recursive star definition in quantales}
\label{s:star-quantale}

In our final technical section we contextualise our previous
constructions for Conway semirings and variants of Kleene algebras,
showing that the recursive definition \eqref{eq:star} of the Kleene
star is derivable in convolution algebras formed by an arbitrary
catoid $C$ and a quantale $Q$, which we fix for this section. In
convolution quantales $Q^C$, by definition of the star in quantales,
$f^\ast = \bigvee_{i\ge 0} f^i$ with $f^0$ = $\id_0$ and
$f^{i+1} = f\ast f^i$. So $f^\ast(x)=\bigvee_{i\ge 0}f^i(x)$ can be
used in proofs.

  First we adapt \autoref{lem:conv-star-prop} from dioids to quantales.

\begin{lem}
  \label{lem:conv-star-prop-quantale}
  Let $C$ be a catoid and $Q$ a quantale. 
  Then for all $f:C\to Q$,
  $x\in C_1$ and $n\ge 1$,
  \begin{equation}
    \label{eq:power-exact-quantale}
    f^n(x) = \bigvee_{y,z\in C} \bigvee_{i=0}^{n-1} f^i(s(x))
    f(y)f^{n-1-i}(z)[x\in y\odot z,z\neq s(x)].
  \end{equation}
\end{lem}
\begin{proof}
Replace finite sums by arbitrary sups in the proof of \autoref{lem:conv-star-prop}.
\end{proof}

\begin{prop}
  \label{prop:conv-star}
  For all $f:C\to Q$, $e\in C_0$ and $x\in C_1$, the equations in
  \eqref{eq:star} are derivable:
 \begin{equation*}
   f^\ast(e)= f(e)^\ast,\qquad  f^\ast(x) = f^\ast(s(x))\bigvee_{y,z\in C} f(y)f^\ast(z)[x\in
   y\odot z,y\neq s(x)].
\end{equation*}
\end{prop}
\begin{proof}
  For the first equation, we first show $f^n(e)=f(e)^n$ by a
  straightforward induction on $n$. For $n=0$,
  \begin{equation*}
    f^0(e)=id_0(e)=1 = f(e)^0,
  \end{equation*}
  For $n+1$ we have 
  \begin{equation*}
    f^{n+1}(e)= (f\ast f^n)(e)= f(e)f^n(e)= f(e)f(e)^n= f(e)^{n+1}
  \end{equation*}
  and therefore
  \begin{equation*}
    f^\ast(e)=\bigvee_if^i(e)= \bigvee_i f(e)^i = f(e)^\ast.
  \end{equation*}

  For the second equation we abbreviate again
  $I(x,y,z)=[x\in y\odot z,y\neq s(x)]$. Then
  \begin{align*}
    f^\ast(x)
    & = \id_0(x) \lor \bigvee_{n\ge 1} f^n(x)\\
    &= \bigvee_{n\ge 1} \bigvee_{y,z\in C} \bigvee_{i=0}^{n-1} f^i(s(x))
      f(y)f^{n-1-i}(z)I(x,y,z)\\
    &= \bigvee_{y,z\in C} \bigvee_{n\ge 1}\bigvee_{i=0}^{n-1} f^i(s(x))
      f(y)f^{n-1-i}(z)I(x,y,z)\\
    &= \bigvee_{y,z\in C} \bigvee_{i\ge 0}\bigvee_{n>i} f^i(s(x))
      f(y)f^{n-1-i}(z)I(x,y,z)\\
    &= \bigvee_{y,z\in C} \bigvee_{i\ge 0}\bigvee_{m\ge 0} f^i(s(x))
      f(y)f^m(z)I(x,y,z)\\
    &= \bigvee_{y,z\in C} \left(\bigvee_{i\ge 0}f^i(s(x))\right)
      f(y)\left(\bigvee_{m\ge 0}  f^m(z)\right)I(x,y,z)\\
    &= \bigvee_{y,z\in C} f^\ast(s(x))
      f(y) f^\ast(z)I(x,y,z)\\
    &= f^\ast(s(x))\bigvee_{y,z\in C}
      f(y) f^\ast(z)I(x,y,z).
  \end{align*}
In the second step, $\id_0(x)=\bot$ because $x\in C_1$ and we use
\eqref{eq:power-exact-quantale} to expand $f^n$. In the fourth step we use the
constraint $0\le i<n$ on the variable range to swap the sums on $i$
and $n$. In the fifth we set $m=n-1-i$.
\end{proof}

Note the similarity between the proofs of \autoref{thm:star-cont-ka-star}
and \autoref{prop:conv-star}.

%%%%%%%%%%%%%%%%%%%%%%%%%%%%%%%%%%%%%%%%%%%%%%%%%%%%%%

\section{Conclusion}\label{s:conclusion}

We have generalised a previous construction of convolution quantales
from categories and catoids to convolution Kleene algebras and
convolution Conway semirings, using Möbius catoids to restrict the
infinite sups in convolution quantales to finite ones, and making
Kuich and Salomaa's classical recursive construction of the Kleene
star on the convolution algebra available in a much more general
setting. Möbius catoids provide precisely the concepts needed: a
notion of length on elements of the underlying catoids supports a
recursive definition and the verification of the convolution algebra
axioms by induction, while a notion of finite decomposability of
elements allows taking finite sups or sums of them. Our main technical
result, \autoref{thm:main-theorem}, allows constructing convolution
Kleene algebras on Möbius catoids in various contexts, from
convolution Kleene algebras with tests and modal convolution Kleene
algebras to higher convolution Kleene algebras, in particular
concurrent convolution Kleene algebras, and for a wide range of models
and applications.

Convolution Kleene algebras and quantales complement each other. The
latter work on larger classes of categories or catoids, capturing for
instance general path problems, and allowing more general types of
domain and codomain operations on convolution algebras. A Kleene star
can be defined simply as a sum of powers and in particular for models
like weighted relations and matrix algebras, where the Möbius
conditions do not apply. Semiring-based approaches such as Kleene
algebras, by contrast, are more general and appear in many computing
applications; they are closer to program semantics and characterise
models defined in terms of generators and relations more succinctly,
for instance formal power series on words, path algebras generated by
finite graphs, programs generated by atomic commands and tests, or
more general path algebras such as polygraphs in higher-dimensional
rewriting. It can also be expected that convolution Kleene algebras
have more appealing completeness, decidability and complexity
properties than their quantalic companions.

Our aim in this article lies in the foundations of convolution Kleene
algebras on Möbius catoids and Möbius categories. As stepping stones
towards programming applications, we envisage concrete quantitative
Hoare logics, predicate transformer semantics or interval temporal
logics, with weighted or probabilistic ``predicates'' or programs,
including distributed or concurrent ones. Applications in
higher-dimensional rewriting might include, for instance, rewriting
using labels such as in decreasing diagram techniques~\cite{Oostrom94}
or probabilistic and weighted approaches generalising those for
classical rewriting
systems~\cite{BournezK02,Faggian22,GavazzoF23}. More generally, it
seems interesting to consider the higher-dimensional ``homotopical''
aspects of our approach in contexts such as games, networks and
distributed systems, where groups of agents interact, cooperate or
exchange information.

%%%%%%%%%%%%%%%%%%%%%%%%%%%%%%%%%%%%%%%%%%%%%%%%%%%%

\section*{Acknowledgement}
  \noindent J.~Wagemaker was partially supported by the NWO under grant no.\ VI.Veni.242.134 (VerHyp).

\bibliographystyle{alphaurl}
\bibliography{conv-ka}

\appendix

\section{Independence of closure axioms in \texorpdfstring{$n$}{n}-semirings}
\label{A:independence}

Here we show that in $n$-semirings, the two closure axioms for domain
and codomain are independent. There are $n$-semirings which satisfy
the other $n$-semiring axioms but not the closure axioms, and there
are $n$-semirings which satisfy all $n$-semiring axioms except the
codomain closure axiom.

First consider the $2$-fold modal semiring $0 < 1_0 < 1_1 < a$ with
\begin{equation*}
  \begin{array}{c|cccc}
    \cdot_0 & 0 & 1_0 & 1_1 & a\\
    \hline
    0 & 0 & 0& 0 & 0\\
    1_0 & 0 & 1_0 & 1_1 & a\\
    1_1 & 0 & 1_1 & a & a\\
    a & 0 & a & a & a
  \end{array}
  \qquad\qquad
   \begin{array}{c|cccc}
    \cdot_1 & 0 & 1_0 & 1_1 & a\\
    \hline
     0 & 0 & 0 & 0 & 0 \\
    1_0 & 0 & 1_0 & 1_0 & 1_0\\
     1_1 & 0 & 1_0 & 1_1 & a\\
    a & a & 0 & a & a
   \end{array}
   \qquad\qquad
   \begin{array}{c|c|c}
     & \dom_0  = \cod_0 & \dom_1 = \cod_1\\
     \hline
    0 & 0 & 0 \\
    1_0 & 1_0 & 1_0 \\
    1_1 & 1_0& 1_1\\
    a & 1_0 & 1_1
  \end{array}
\end{equation*}
It satisfies all the $n$-semiring axioms except the two closure axioms
in the last line, which fail:
\begin{equation*}
  \dom_1(\dom_1(1_1)\cdot_0 \dom_1(1_1)) = \dom_1(1_1 \cdot_0 1_1)
  = \dom_1(a)= 1_1 < a = \dom_1(1_1)\cdot_0 \dom_1(1_1),
\end{equation*}
and likewise
$\cod_1(\cod_1(1_1)\cdot_0 \cod_1(1_1)) = 1_1 < a = \cod_1(1_1)\cdot_0
\cod_1(1_1)$.

Second, consider the $2$-fold modal semiring $0  < 1_0 <  a,1_1 <
b$ $10 = a2$, $11=a_5$, with

\begin{equation*}
  \begin{array}{c|ccccc}
    \cdot_0 & 0 & 1_0 & a& 1_1 & b\\
    \hline
    0 & 0 & 0 & 0& 0 & 0\\
    1_0 & 0 & 1_0& a & 1_1 & b\\
     a & 0 & a & a & b & b\\
     1_1 & 0 & 1_1 & b & b & b\\
     a_1 & 0 & b & b & b & b
  \end{array}
\qquad\qquad
 \begin{array}{c|ccccc}
    \cdot_1 & 0 & 1_0 & a & 1_1 & b\\
    \hline
    0 & 0 & 0 & 0& 0 & 0\\
    1_0 & 0 & 1_0 & a & 1_0 & a\\
     a & 0& 1_0 & a & a & a\\
     1_1 & 0 & 1_0 & a & 1_1 & b\\
     b & 0 & 1_0 & a & b & b
 \end{array}
 \qquad\qquad
 \begin{array}{c|c|c|c}
   & \dom_0 = \cod_0 & \dom_1 & \cod_1\\
   \hline
   0 & 0 & 0 & 0\\
   1_0 & 1_0 & 1_0 & 1_0\\
   a & 1_0 & 1_0& 1_1\\
   1_1 & 1_0 & 1_1 & 1_1\\
   b & 1_0 & 1_1 & 1_1
    \end{array}
  \end{equation*}
  It satisfies all $n$-semiring axioms except $\cod_1$-closure,
  because
  \begin{equation*}
    \cod_1(\cod_1(1_1)\cdot_0 \cod_1(a)) = \cod_1(1_1\cdot_0 1_1) =
    \cod_1(b) = 1_1 < b < cod_1(1_1)\cdot_0 \cod_1(a)).
  \end{equation*}

\end{document}